# An Optimal and Progressive Approach to Online Search of Top-K Influential Communities

Fei Bi[†], Lijun Chang[§] ✉, Xuemin Lin[†], Wenjie Zhang[†]
[†] University of New South Wales, Australia [§] The University of Sydney, Australia
f.bi@student.unsw.edu.au, Lijun.Chang@sydney.edu.au {lxue,zhangw}@cse.unsw.edu.au

## Abstract

Community search over large graphs is a fundamental problem in graph analysis. Recent studies propose to compute top-$k$ influential communities, where each reported community not only is a cohesive subgraph but also has a high influence value. The existing approaches to the problem of top-$k$ influential community search can be categorized as index-based algorithms and online search algorithms without indexes. The index-based algorithms, although being very efficient in conducting community searches, need to pre-compute a special-purpose index and only work for one built-in vertex weight vector. In this paper, we investigate online search approaches and propose an *instance-optimal* algorithm LocalSearch whose time complexity is linearly proportional to the size of the smallest subgraph that a correct algorithm needs to access without indexes. In addition, we also propose techniques to make LocalSearch *progressively* compute and report the communities in decreasing influence value order such that $k$ does not need to be specified. Moreover, we extend our framework to the general case of top-$k$ influential community search regarding other cohesiveness measures. Extensive empirical studies on real graphs demonstrate that our algorithms outperform the existing online search algorithms by several orders of magnitude.

## 1 Introduction

Community search is a fundamental problem in graph analysis, and has been receiving increasing interest in recent years (see a recent tutorial in [22] and references therein). Existing works on community search mainly focus on the cohesiveness of structural connections among members of a community while ignoring other aspects of communities, *e.g.*, influence. As a result, an enormous number of overlapping communities may be reported, and also a single community can be of a large size. However, in many application domains, we are usually only interested in the most influential communities [26]. Motivated by this, top-$k$ influential community search is recently proposed and studied in [8, 26, 27]. It has many important applications such as detecting cohesive communities consisting of celebrities or influential people in social networks [25], finding strong collaboration patterns among influential researchers in research-collaboration networks [26, 27], and extracting backbone structures (*i.e.*, being both cohesive and influential) from biology networks [4]. Besides, computing top-$k$ influential communities also greatly refines communities to their core members [26].

Here, the graph $G = (V, E)$ is associated with a vertex weight vector $\omega(\cdot)$ assigning an influence value to every vertex in $V$. Each community of $G$, called *influential $\gamma$-community*, besides being a cohesive subgraph (*i.e.*, with minimum degree at least $\gamma$), has *an influence value that equals the minimum vertex weight of the community* [26]. As a result, members in a high influential $\gamma$-community are highly connected to each other, and moreover each member is also an influential individual. Formally speaking, a connected subgraph $g$ of $G$ is an influential $\gamma$-community [26] if 1) its minimum vertex degree is at least $\gamma$, and 2) it is the maximal one among all such subgraphs of $G$ with the same influence value as $g$. For example, consider the graph in Figure 1 where vertex weights are shown beside the vertices, and $\gamma = 3$. There are two influential $\gamma$-communities: the subgraphs induced by vertices $\{v_0, v_1, v_5, v_6\}$ and vertices $\{v_3, v_4, v_7, v_8, v_9\}$ that, respectively, have influence values 10 and 13. The subgraph induced by vertices $\{v_3, v_4, v_7, v_8\}$ also has an influence value 13; however, it is not an influential $\gamma$-community since it is not maximal. *The problem of top-k influential community search is to compute the k influential $\gamma$-communities of a graph with the highest influence values, for a user-specified query consisting of $\gamma$ and k.*

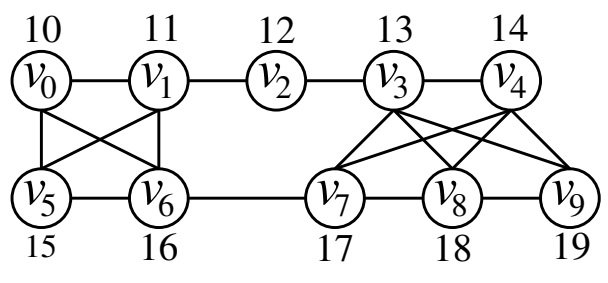

Figure 1: An example graph

**Existing Approaches and Their Deficiencies.** The existing approaches to top-$k$ influential community search can be categorized as *index-based algorithms* and *online search algorithms*.

<u>*(1) Index-based Algorithms.*</u> Li et al. [26] proposed an index-based algorithm IndexAll to efficiently retrieve the top-$k$ influential $\gamma$-communities from a pre-built special-purpose index that essentially materializes all influential $\gamma$-communities of a graph in a compact form for all possible $\gamma$ values. However, the special-purpose index adds a large burden to the graph processing system, as it is time-consuming to update the index when the graph changes. Moreover, IndexAll cannot process queries that impose vertex weight vectors different from the one used in the index.

*(2) Online Search Algorithms.* Online search algorithms without pre-computing indexes are investigated in [8, 26]. Firstly, Li et al. [26] proposed an OnlineAll algorithm, which online computes all influential $\gamma$-communities in a graph in increasing influence value order. OnlineAll iteratively applies the following three subroutines: 1) reduce the current graph to its $\gamma$-core (*i.e.*, maximal subgraph with minimum degree at least $\gamma$); 2) identify the connected component of the resulting graph containing the vertex with the minimum weight, which is the next influential $\gamma$-community; and 3) remove the minimum-weight vertex from the graph. During this process, the last $k$ identified influential $\gamma$-communities are the results. Among the above three subroutines of OnlineAll, the second one is the most time-consuming due to the overlapping nature of the influential $\gamma$-communities [8]. In view of this, Chen et al. [8] proposed a Forward algorithm which conducts the second subroutine of OnlineAll (*i.e.*, connected component computation) only for the last $k$ iterations; as a result, Forward improves upon OnlineAll. Nevertheless, both OnlineAll and Forward are global search algorithms that need to traverse the entire graph for finding just the top-$k$ influential $\gamma$-communities.

**Challenges and Our Online Local Search Approach.** In this paper, we aim to compute the top-$k$ influential $\gamma$-communities by conducting a *local search* on the graph $G$ *without pre-computing indexes*, to overcome the deficiencies of the existing algorithms. The benefits of local search without indexes are two-fold.

- It does not incur any burden to the graph data management system, regarding index construction and index maintenance.
- It can efficiently process a query by visiting only a small portion of the graph $G$.

However, there are three challenges to tackle to achieve this.

- It is challenging to determine whether a given subgraph of $G$ is sufficient for processing a query.
- It is challenging to choose a proper subgraph to process.
- It is challenging to carry out the ideas efficiently for real-time query processing over large graphs.

Note that, the Backward algorithm proposed in [8] tried to conduct a local search for computing top-$k$ influential $\gamma$-communities, but it fails by having a quadratic time complexity and is outperformed by Forward when $\gamma$ is large [8].

We propose a local search framework to tackle the above challenges, based on the following ideas. Firstly, we prove that *if the subgraph $G_{\geq\tau}$ of $G$ contains at least $k$ influential $\gamma$-communities, then the top-$k$ influential $\gamma$-communities in $G_{\geq\tau}$ is the query result*, where $G_{\geq\tau}$ denotes the subgraph of $G$ induced by all vertices with weights at least $\tau$. Thus, our goal is to find the smallest subgraph $G_{\geq\tau^*}$ of $G$ containing at least $k$ influential $\gamma$-communities. Secondly, we prove that *the number of influential $\gamma$-communities in a subgraph $G_{\geq\tau}$ of $G$ is non-decreasing when $\tau$ decreases*. Thus, we can find the target subgraph $G_{\geq\tau^*}$ by iteratively decreasing the value of $\tau$ until reaching the target value. Thirdly, to efficiently implement the above ideas, we propose to only process (*i.e.*, count the number of influential $\gamma$-communities for) the subgraphs $G_{\geq\tau_1}, G_{\geq\tau_2}, \ldots$, such that the size of $G_{\geq\tau_i}$ is around twice the size of $G_{\geq\tau_{i-1}}$ for every $i > 1$. For example, to compute the top-2 influential $\gamma$-communities in the graph in Figure 2(a) with $\gamma = 3$, we first count the number of influential $\gamma$-communities in the subgraph $G_{\geq 9}$ as shown in Figure 2(b), which is 1. Thus, we need to find another smaller $\tau$ such that the size of $G_{\geq\tau}$ is around twice the size of $G_{\geq 9}$; we obtain $\tau_2 = 5$ and $G_{\geq 5}$ is shown in Figure 2(c). As there are three influential $\gamma$-communities in $G_{\geq 5}$ — the subgraphs induced by vertices $\{v_0, v_1, v_5, v_6\}$, $\{v_3, v_4, v_8, v_9\}$ and $\{v_3, v_4, v_8, v_9, v_{10}\}$, respectively — the top-2 are the result.

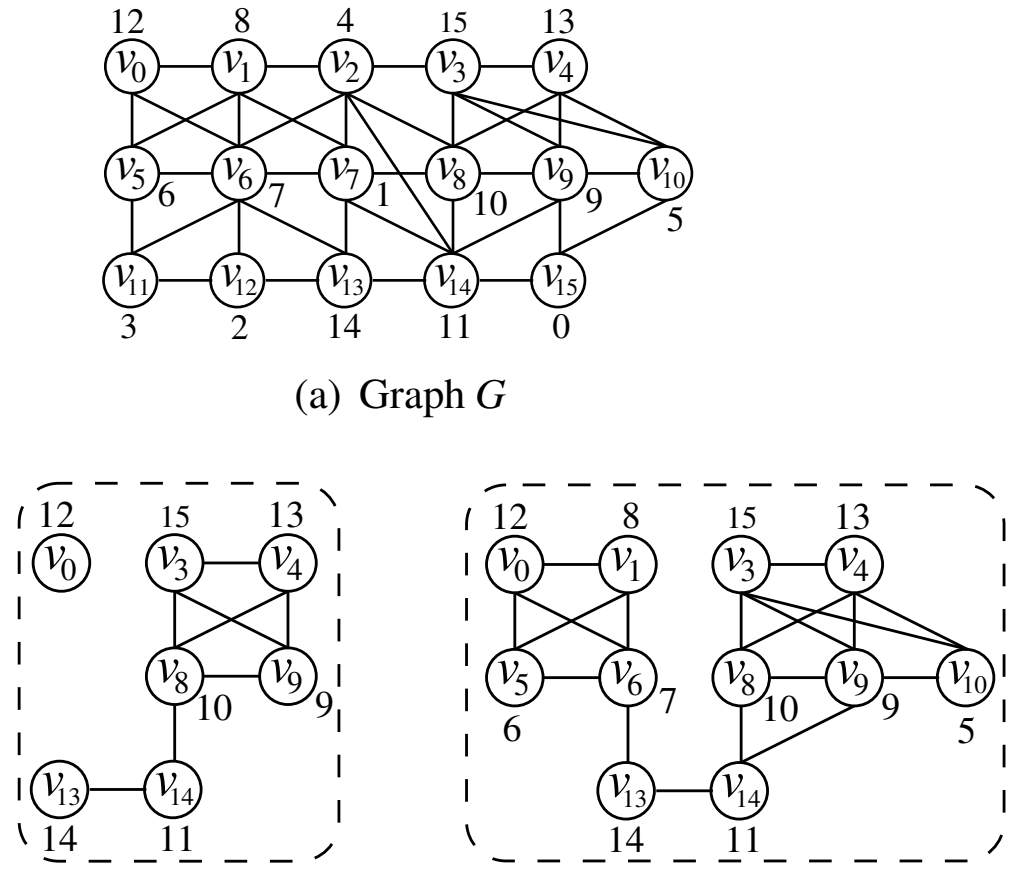

(a) Graph $G$

(b) Subgraph $G_{\geq 9}$

(c) Subgraph $G_{\geq 5}$

Figure 2: An example of our local search framework

As a critical subroutine in our local search framework, we propose a linear-time algorithm to count the number of influential $\gamma$-communities in an arbitrary given subgraph of the graph $G$. As a result, we prove that the time complexity of our local search algorithm LocalSearch is linear to the size of the largest subgraph that it accesses. We also show that the subgraph that LocalSearch accesses is at most a constant (specifically, 3) times larger than the smallest subgraph $G_{\geq\tau^*}$ that an online search algorithm without indexes needs to access for correctly computing the top-$k$ influential $\gamma$-communities. Thus, LocalSearch is instance-optimal among the class of online search algorithms without indexes.

Moreover, we propose techniques to make LocalSearch progressively compute and report the influential $\gamma$-communities in decreasing influence value order such that $k$ does not need to be specified in the query. The user can terminate the algorithm at any time once determining that enough influential $\gamma$-communities have been reported. Our instance-optimality result of LocalSearch also carries over to the progressive approach. It is worth noting that the existing global search algorithms OnlineAll and Forward are only able to report the $k$ communities at the end of the algorithm.

Finally, we also extend our local search framework to the case of non-containment community search and to the case of top-$k$ influential community search regarding other cohesive-ness measures.

**Contributions.** Our main contributions are summarized as follows.

- We propose an instance-optimal algorithm LocalSearch, whose time complexity is linearly proportional to the size of the smallest subgraph that a correct algorithm without indexes needs to access, for computing the top-$k$ influential $\gamma$-communities (Section 3).
- We propose techniques to make LocalSearch progressively compute and report the influential $\gamma$-communities in decreasing influence value order (Section 4).
- We extend our local search framework to the general case of top-$k$ influential community search regarding other cohesiveness measures (Section 5).

Extensive experimental results in Section 6 show that our local search algorithms outperform the existing online search algorithms by several orders of magnitude.

**Related Works.** Besides top-$k$ influential community search as discussed above, other related works are categorized as follows.

*(1) Community Detection.* Community detection is a long-studied problem [15], which aims to find all communities in a graph for a given community definition. A community is a group of vertices that are similar to each other and dissimilar to vertices outside the community. The existing community definitions can be categorized as, *(1) graph partitioning* that divides the vertices of a graph into $k$ groups of predefined size such that the number of inter-group edges is minimized [2, 24, 37], *(2) hierarchical clustering* that reveals the multi-level structure of the graph by computing the similarity for each pair of vertices [18, 28], *(3) partitional clustering* that divides vertices into $k$ clusters such that the cost function defined on distances/disimilarities between vertices is minimized [19, 30, 33], and *(4) spectral clustering* that partitions the graph by using the eigenvectors of the matrix derived from the graph [14, 29, 35]. Due to inherent problem natures, these techniques cannot be used to compute top-$k$ influential communities studied in this paper.

*(2) Cohesive Subgraph Computation.* Computing cohesive subgraphs in a graph has been extensively studied in [6, 7, 11, 17, 31, 32, 34, 40], where a cohesive subgraph can be regarded as a community. The cohesiveness of a graph is measured by the minimum degree (aka, $k$-core) [32, 34], the average degree (aka, edge density) [7, 17], the minimum number of triangles each edge participates in (aka, $k$-truss) [11, 31], or the edge connectivity (aka, $k$-edge connected components) [6, 40]. These works focus on computing all maximal subgraphs whose cohesiveness is no smaller than a user-given threshold. Due to different problem definitions, these techniques cannot be applied to the problem studied in this paper.

*(3) Community Search.* Recently, cohesive community search is receiving increasing interests (see [22] and references therein). Given one query vertex or a set of query vertices, cohesive community search is to find a subgraph such that (1) it contains all query vertices and (2) its cohesiveness is no smaller than the user given threshold. For example, $k$-core-based community search is studied in [1, 16, 36], edge density-based community search is studied in [39], $k$-truss-based community search is studied in [21, 23], and edge connectivity-based community search is studied in [5, 20]. As influences of vertices are not considered in these works, these techniques cannot be applied to the problem of top-$k$ influential community search.

# 2 Preliminaries

In this paper, we focus on a *vertex-weighted undirected graph* $G = (V, E, \omega)$, where $V$ is the set of vertices, $E \subseteq V \times V$ is the set of edges, and $\omega$ is a weight vector that assigns each vertex $u \in V$ a weight denoted by $\omega(u)$. Here, the weight $\omega(u)$ represents the *influence* of vertex $u$, which can be its PageRank value, centrality score, h-index, social status, and etc; the larger the value, the more influential the vertex is. Following the existing works [8, 26], we assume that the weights of vertices are pre-given[1], and each vertex has a distinct weight (*i.e.*, $\omega(u) \neq \omega(v), \forall u \neq v$). For a given value $\tau$, we use $V_{\geq\tau}$ to denote the subset of $V$ consisting of all vertices with weights no less than $\tau$ (*i.e.*, $V_{\geq\tau} = \{u \in V \mid \omega(u) \geq \tau\}$). In the following, for ease of presentation we simply refer to a vertex-weighted undirected graph as a graph when the context is clear.

We denote the size of a graph $G$ by $\texttt{size}(G)$, which is the summation of the number of vertices and the number of edges in $G$; that is, $\texttt{size}(G) = |V| + |E|$. The set of neighbors of $u \in V$ in $G$ is denoted by $N(u) = \{v \in V \mid (u, v) \in E\}$, and the degree of $u$ is denoted by $d(u)$, which is the number of neighbors of $u$ (*i.e.*, $d(u) = |N(u)|$). Given a subset $S \subseteq V$ of vertices, the subgraph of $G$ induced by $S$ is denoted by $G[S]$, which consists of all edges of $G$ whose both end-points are in $S$; that is, $G[S] = (S, \{(u, v) \in E \mid u, v \in S\}, \omega)$. For presentation simplicity, we use $G_{\geq\tau}$ to denote the subgraph of $G$ induced by vertices $V_{\geq\tau}$ (*i.e.*, $G_{\geq\tau} = G[V_{\geq\tau}]$).

**Influential Community.** This paper aims to identify influential communities from a given large graph $G$, where each community is a cohesive subgraph of $G$ and has an influence value. The influence value of a subgraph is defined in below, which is shown to be robust to outliers as discussed in [26].

Definition 2.1: [26] Given a subgraph $g = (V(g), E(g), \omega)$ of $G$, the ***influence value*** of $g$, denoted by $f(g)$, is defined as the minimum weight of the vertices in $g$ (*i.e.*, $f(g) = \min_{u \in V(g)} \omega(u)$).

For the cohesiveness measure, many definitions have been proposed and studied in the literature, *e.g.*, $k$-core [32, 34], edge density [7, 17], $k$-truss [11, 31], edge connectivity [6, 10, 40]. Among them, the $k$-core-based cohesiveness measure has been widely adopted, due to its simplicity and fast computability. Thus, we mainly focus on $k$-core-based community search in the following, and will extend our techniques to other cohesiveness measures in Section 5.

[1] Note that, the techniques proposed in this paper can be extended to the case that the weights of vertices are computed online based on a query, *e.g.*, the weight of a vertex is the reciprocal of the shortest distance to query vertices as studied in closest community search [23]. We will analyze the time complexities of such extensions in our future work.

Definition 2.2: [26] Given a graph $G$ and an integer $\gamma$, an ***influential $\gamma$-community*** is a *vertex-induced subgraph* $g$ of $G$ such that the following constraints are satisfied.

- ***Connected:*** $g$ is a connected subgraph;
- ***Cohesive:*** each vertex $u$ in $g$ has a degree at least $\gamma$, *i.e.*, the minimum degree of $g$ is at least $\gamma$;
- ***Maximal:*** there exists no other subgraph $g'$ of $G$ such that (1) $g'$ is a supergraph of $g$ with $f(g') = f(g)$, and (2) $g'$ is also connected and cohesive.

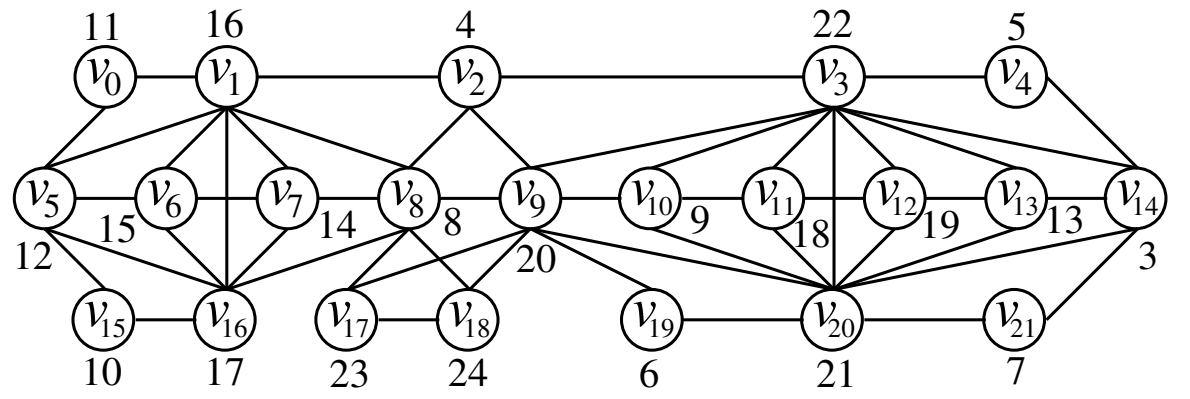

Figure 3: A graph

Example 2.1: Consider the graph in Figure 3 and $\gamma = 3$. The subgraph $g_1$ induced by vertices $\{v_3, v_{10}, v_{11}, v_{12}, v_{20}\}$ is connected, and has a minimum degree 3 and an influence value 9. However, it is not an influential $\gamma$-community because it is not maximal; that is, the subgraph $g_2$ induced by vertices $\{v_3, v_9, v_{10}, v_{11}, v_{12}, v_{13}, v_{20}\}$ is an influential $\gamma$-community with the same influence value as $g_1$. Note that, the subgraph induced by vertices $\{v_3, v_{11}, v_{12}, v_{20}\}$ is also an influential $\gamma$-community; this is because, although it is a subgraph of $g_2$, it has a larger influence value (*i.e.*, 18) than $g_2$. □

In the following, for presentation simplicity, we simply refer to an influential $\gamma$-community by the set of vertices from which the influential $\gamma$-community is induced.

**Problem Statement.** Given a graph $G = (V, E, \omega)$, and two query parameters $\gamma$ and $k$, the problem of top-$k$ influential community search is to extract the $k$ influential $\gamma$-communities with the highest influence values from $G$.

For example, consider the graph in Figure 3 with $\gamma = 3$ and $k = 4$. The top-4 influential $\gamma$-communities are $\{v_3, v_{11}, v_{12}, v_{20}\}$, $\{v_1, v_6, v_7, v_{16}\}$, $\{v_3, v_{11}, v_{12}, v_{13}, v_{20}\}$ and $\{v_1, v_5, v_6, v_7, v_{16}\}$ with influence values 18, 14, 13 and 12, respectively.

# 3 A Local Search Approach

In the following, we first develop a local search framework for efficient top-$k$ influential community search in Section 3.1, and then present our approach in Section 3.2, while the instance-optimality of our local search approach is illustrated in Section 3.3.

## 3.1 The Framework

**Properties of Influential $\gamma$-community.** Firstly, we prove some important properties of influential $\gamma$-community in the following lemmas and theorems.

Lemma 3.1: *For any two values $\tau_1 \le \tau_2$, every influential $\gamma$-community in $G_{\ge\tau_2}$ is also an influential $\gamma$-community in $G_{\ge\tau_1}$. Note that, $G_{\ge\tau_1}$ is a supergraph of $G_{\ge\tau_2}$.*

Proof: We prove the lemma by contradiction. Assume there is an influential $\gamma$-community $g$ in $G_{\ge\tau_2}$ that is not an influential $\gamma$-community in $G_{\ge\tau_1}$. Note that, $g$ is also a subgraph of $G_{\ge\tau_1}$. Then, $g$ must violate the maximality constraint of influential $\gamma$-community in $G_{\ge\tau_1}$, since the connectivity and the cohesiveness constraints are satisfied for $g$. That is, there is a subgraph $g'$ of $G_{\ge\tau_1}$ that (1) is a supergraph of $g$ with $f(g') = f(g)$, and (2) also satisfies the connectivity and cohesiveness constraints. From the definition of influence value of a subgraph and the fact that $g$ is a subgraph of $G_{\ge\tau_2}$, $g'$ must also be a subgraph of $G_{\ge\tau_2}$; this contradicts that $g$ is an influential $\gamma$-community in $G_{\ge\tau_2}$. Thus, the lemma holds. □

Lemma 3.2: *For any two values $\tau_1 \le \tau_2$ and an influential $\gamma$-community $g$ in $G_{\ge\tau_1}$, if the influence value of $g$ is no smaller than $\tau_2$, then $g$ is also an influential $\gamma$-community in $G_{\ge\tau_2}$.*

Proof: It is easy to see that for such an influential $\gamma$-community $g$ with $f(g) \ge \tau_2$, $g$ is a subgraph of $G_{\ge\tau_2}$ and it satisfies the connectivity and the cohesiveness constraints. Moreover, $g$ is maximal in $G_{\ge\tau_2}$ since (1) it is maximal in $G_{\ge\tau_1}$ and (2) $G_{\ge\tau_2}$ is a subgraph of $G_{\ge\tau_1}$. Thus, $g$ is an influential $\gamma$-community in $G_{\ge\tau_2}$. □

Theorem 3.1: *Let $\tau^*$ be the largest value such that $G_{\ge\tau^*}$ contains at least k influential $\gamma$-communities. Then, the set of top-k influential $\gamma$-communities in $G_{\ge\tau^*}$ is the set of top-k influential $\gamma$-communities in G.*

Proof: First of all, we assume that $G$ contains at least $k$ influential $\gamma$-communities; otherwise, $\tau^*$ in the statement of the theorem is not properly defined. Let $\tau_{min}$ be the minimum vertex weight in $G$, then $G_{\ge\tau_{min}}$ is the same as $G$ and moreover $\tau_{min} \le \tau^*$. From Lemma 3.1, we know that each influential $\gamma$-community in $G_{\ge\tau^*}$ is also an influential $\gamma$-community in $G$. It is easy to see that all influential $\gamma$-communities in $G_{\ge\tau^*}$ have influence values at least $\tau^*$. From Lemma 3.2, we also know that each influential $\gamma$-community in $G$ that is not contained in $G_{\ge\tau^*}$ must have an influence value smaller than $\tau^*$. Thus, the theorem holds. □

**The Framework.** Following Theorem 3.1, to compute the top-$k$ influential $\gamma$-communities in $G$, we can first identify the largest influence value $\tau^*$ such that $G_{\ge\tau^*}$ contains at least $k$ influential $\gamma$-communities, and then return the set of top-$k$ influential $\gamma$-communities in $G_{\ge\tau^*}$ as the result. In this way, we only need to work on the subgraph $G_{\ge\tau^*}$ which can be much smaller than $G$. For example, $\frac{\texttt{size}(G_{\ge\tau^*})}{\texttt{size}(G)}$ is smaller than 0.073% across all the graphs tested in our experiments for $k = 10$ and $\gamma = 10$. However, it is non-trivial to obtain the appropriate influence value $\tau^*$.

From Lemma 3.1, we know that the number of influential $\gamma$-communities in the subgraph $G_{\ge\tau}$ increases along with the decreasing of $\tau$. Thus, one possible way to computing $\tau^*$ is conducting a binary search on the sequence of all possible vertex weights in $G$. However, it is time consuming to count the number of influential $\gamma$-communities in a graph, which takes linear time to the size of the graph (see Section 3.2.1), and the size of the first subgraph of $G$ tested by the binary search may be as large as half of $\texttt{size}(G)$. Thus, binary search does not save the computational cost.

**Algorithm 1:** LocalSearch

**Input**: A graph $G = (V, E, \omega)$, and two integers $k$ and $\gamma$
**Output**: Top-$k$ influential $\gamma$-communities in $G$

```
1 τ1 ← the largest τ value such that G≥τ would contain at least k
    influential γ-communities;
2 i ← 1;
3 while CountIC(G≥τi, γ) < k and G≥τi ≠ G do
4     τi+1 ← max {{τ | size(G≥τ) ≥ δ · size(G≥τi)} ∪ {τmin}};
      /* τmin is the smallest vertex weight in G */;
5     i ← i + 1;
6 return top-k communities in EnumIC(G≥τi);
```

In this paper, we propose to use the exponential growth strategy for computing the target $\tau$ value; that is, we iteratively increase the size of the graph $G_{\geq\tau}$, with a growing ratio of $\delta$, for processing. The proper setting of $\delta$ will be discussed in Section 3.3. The pseudocode of our framework is shown in Algorithm 1. We first heuristically compute the largest $\tau_1$ value such that $G_{\geq\tau_1}$ would contain at least $k$ influential $\gamma$-communities (Line 1). For example, $\tau_1$ could be set as the $(k+\gamma)$-th largest vertex weight in $G$; that is, the $k$ influential $\gamma$-communities contain at least $k+\gamma$ distinct vertices. Then, as long as $G_{\geq\tau_i}$ contains less than $k$ influential $\gamma$-communities (*i.e.*, $\mathsf{CountIC}(G_{\geq\tau_i}) < k$) and $G_{\geq\tau_i}$ is not the same as $G$ (Line 3), we find the next largest $\tau_{i+1}$ value such that the size of $G_{\geq\tau_{i+1}}$ is at least $\delta$ times the size of $G_{\geq\tau_i}$ (Line 4), and increment $i$ by 1 (Line 5); note that, if $\texttt{size}(G)$ is smaller than $\delta \cdot \texttt{size}(G_{\geq\tau_i})$, then we set $\tau_{i+1}$ as the smallest vertex weight $\tau_{min}$ in $G$. Finally, we compute and return the top-$k$ influential $\gamma$-communities in $G_{\geq\tau_i}$, which is obtained by invoking $\mathsf{EnumIC}(G_{\geq\tau_i})$, as the result (Line 6).

**Graph Organization.** As we will show in Section 3.2.1 that computing the number of influential $\gamma$-communities in a graph $g$ (*i.e.*, $\mathsf{CountIC}(g)$) can be conducted in linear time to the size of $g$, which is measured by the number of vertices and the number of edges in it. Consequently, we also need efficient techniques to retrieve the induced subgraph $G_{\geq\tau}$ in linear time to its size. To do so,

★★ we assume the vertices of $G$ are *pre-sorted* in decreasing order with respect to their weights.

Thus, regarding a $\tau$, the subset $V_{\geq\tau}$ of vertices can be trivially retrieved in $O(|V_{\geq\tau}|)$ time. To also retrieve the induced edges in $G_{\geq\tau}$ in linear time,

★★ we *pre-partition* the adjacent neighbors $N_G(u)$ of each vertex $u$ into two disjoint sets: $N_G^{\geq}(u)$ contains all neighbors of $u$ whose weights are no smaller than $\omega(u)$, and $N_G^{<}(u)$ contains the neighbors of $u$ whose weights are smaller than $\omega(u)$.

These will support efficient online/ad-hoc queries across every $k$ and $\gamma$, while avoiding the maintenance of indexes [26].

| $u$ | $v_{18}$ | $v_{17}$ | $v_3$ | $v_{20}$ | $v_9$ | $v_{12}$ | $v_{11}$ | $v_{16}$ | $v_1$ | $v_6$ | $v_7$ |
|---|---|---|---|---|---|---|---|---|---|---|---|
| $\omega(\cdot)$ | 24 | 23 | 22 | 21 | 20 | 19 | 18 | 17 | 16 | 15 | 14 |
| $u$ | $v_{13}$ | $v_5$ | $v_0$ | $v_{15}$ | $v_{10}$ | $v_8$ | $v_{21}$ | $v_{19}$ | $v_4$ | $v_2$ | $v_{14}$ |
| $\omega(\cdot)$ | 13 | 12 | 11 | 10 | 9 | 8 | 7 | 6 | 5 | 4 | 3 |

(a) Vertices in decreasing weight order

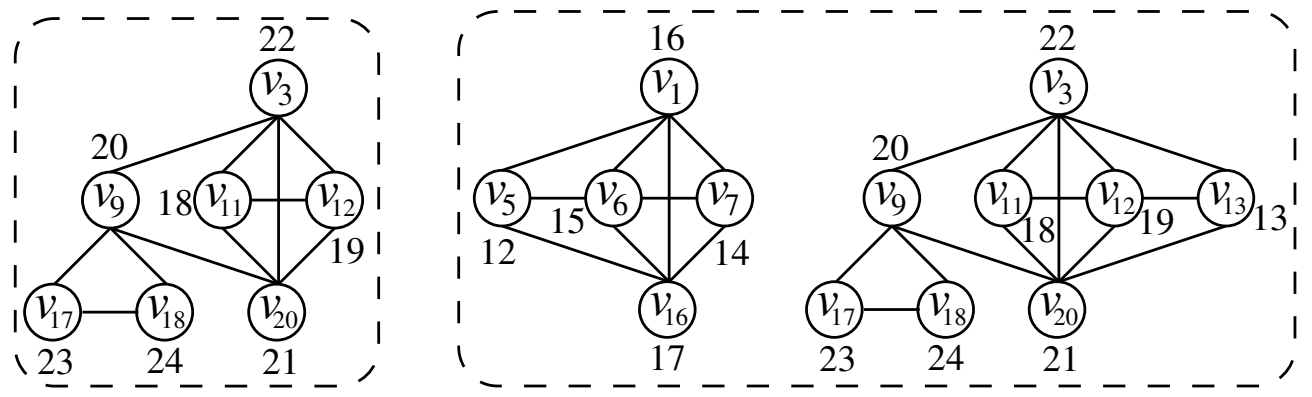

(b) $G_{\geq\tau_1}$ ($\tau_1 = 18$) (c) $G_{\geq\tau_2}$ ($\tau_2 = 12$)

Figure 4: Running example of our local search framework

Thus, to construct $G_{\geq\tau}$, we only need to retrieve the set $N_G^{\geq}(u)$ of neighbors for each $u \in V_{\geq\tau}$, which can be conducted in linear time.

Based on our graph organization, we can efficiently implement Line 4 of Algorithm 1 (*i.e.*, enlarging $G_{\geq\tau_i}$ to obtain $G_{\geq\tau_{i+1}}$ whose size is at least $\delta \cdot \texttt{size}(G_{\geq\tau_i})$) as follows. We first let $G_{\geq\tau_{i+1}}$ be the same as $G_{\geq\tau_i}$, and then iteratively add into $G_{\geq\tau_{i+1}}$ the highest-weighted vertex $u$ in $G \backslash G_{\geq\tau_{i+1}}$ and also an undirected edge between $u$ and each of its neighbors in $N_G^{\geq}(u)$, until the obtained subgraph has a size at least $\delta \cdot \texttt{size}(G_{\geq\tau_i})$. It is easy to see that $G_{\geq\tau_{i+1}}$ is obtained from $G_{\geq\tau_i}$ in time linear to $\texttt{size}(G_{\geq\tau_{i+1}}) - \texttt{size}(G_{\geq\tau_i})$.

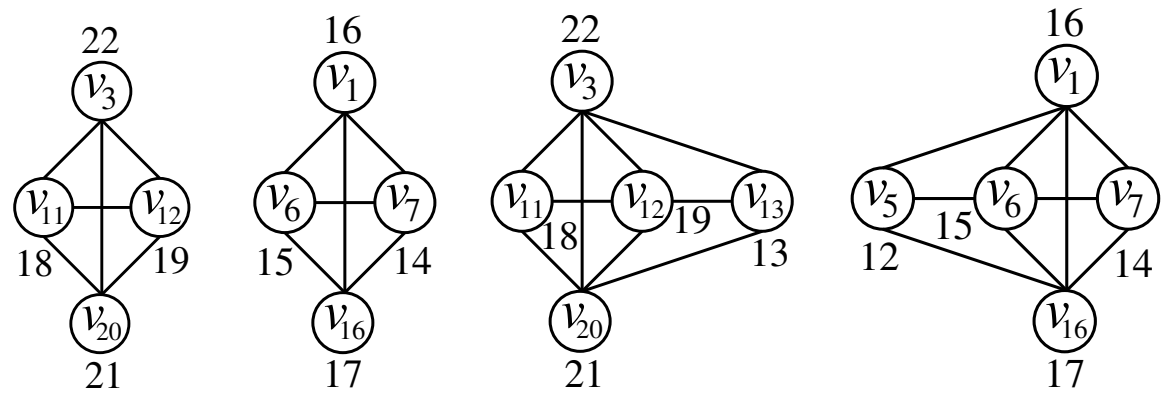

(a) Top-1 (b) Top-2 (c) Top-3 (d) Top-4

Figure 5: Top-4 influential $\gamma$-communities

Example 3.1: Consider the graph $G$ in Figure 3, with $\gamma = 3$ and $k = 4$. The vertices of $G$ in decreasing weight order are shown in Figure 4(a). Initially, we set $\tau_1$ to be the weight of the 7-th vertex (*i.e.*, $v_{11}$) since the top-4 influential $\gamma$-communities will contain at least $k + \gamma = 7$ distinct vertices. Thus, $\tau_1 = 18$ and the subgraph $G_{\geq\tau_1}$ is shown in Figure 4(b). By invoking $\mathsf{CountIC}$ on $G_{\geq\tau_1}$, we know that $G_{\geq\tau_1}$ contains only one influential $\gamma$-community.

Then, we need to find the largest $\tau_2$ value such that the size of $G_{\geq\tau_2}$ is at least $\delta$ times the size of $G_{\geq\tau_1}$; assume $\delta = 2$. As $G_{\geq\tau_1}$ has 7 vertices and 11 edges, the size of $G_{\geq\tau_1}$ is 18. We iteratively add the next highest-weight vertex into the subgraph $G_{\geq\tau_1}$. Firstly, we add $v_{16}$ which has no edges to the subgraph. Secondly, we add $v_1$ with one edge to $v_{16}$ to the subgraph. So on so forth. Until after adding $v_5$ to the subgraph, the size of the subgraph becomes 36. Thus, $\tau_2 = \omega(v_5) = 12$ and $G_{\geq\tau_2}$ is shown in Figure 4(c).

By invoking $\mathsf{CountIC}$ on $G_{\geq\tau_2}$, we know that $G_{\geq\tau_2}$ has four influential $\gamma$-communities. Thus, $\mathsf{EnumIC}$ computes the top-4 influential $\gamma$-communities in $G_{\geq\tau_2}$ as shown in Figure 5, which is outputted as the result of the query. □

**Remark.** In our framework in Algorithm 1, the graph $G$ can be either main memory resident, or disk resident, or stored in a database. The only requirement is that there is an interface to retrieve the vertices (together with their neighbors $N^{\geq}_{G}(\cdot)$) in decreasing weight order. For example, if $G$ is stored on disk, then Algorithm 1 can work in an I/O-efficient manner in a similar way to the semi-external algorithm in [27], as follows. It assumes that the main memory is large enough to store constant information regarding vertices as well as a subset of all edges of $G$, and it sorts edges in decreasing weight order in a preprocessing step, where the weight of an edge equals the minimum weight of its two end-points [27]. Thus, the neighbors in $N^{\geq}_{G}(v)$ of $v$ are stored consecutively on disk, and to construct $G_{\geq \tau_{i+1}}$ from $G_{\geq \tau_i}$, the edges of $G_{\geq \tau_{i+1}}$ that are not in $G_{\geq \tau_i}$ are loaded sequentially from disk to main memory; then, the computations regarding $G_{\geq \tau_{i+1}}$ are conducted in main memory.

In the following, we assume that $G$ is stored in main memory for presentation simplicity; nevertheless, we also evaluate our algorithm for the scenario that $G$ is stored on disk in Section 6.

## 3.2 Our Approach

In Algorithm 1, CountIC can be achieved by invoking EnumIC. However, it is expected that counting the number of influential $\gamma$-communities in a graph would be easier than enumerating them. This is because that, the total size of influential $\gamma$-communities in a graph can be much larger than the size of the graph, since they may overlap with each other [8, 26]. Nevertheless, the existing algorithms do not count the influential $\gamma$-communities in a graph without enumerating them, and they take time at least linear to the size of the top-$k$ influential $\gamma$-communities. Thus, we propose new algorithms for counting, as well as enumerating, influential $\gamma$-communities in a graph in the following two subsections.

### 3.2.1 Influential $\gamma$-community Counting

We first define the notion of keynode regarding influential $\gamma$-community in the following.

Definition 3.1: A vertex $u$ in a graph $G$ is a ***keynode*** regarding a $\gamma$ value if there exists a subgraph $g$ of $G$ such that $g$ has an influence value $\omega(u)$ and the minimum vertex degree of $g$ is at least $\gamma$; note that, this subgraph $g$ must contain $u$ according to the definition of influence value.

For example, $v_7$ in Figure 3 is a keynode regarding $\gamma = 3$, since the subgraph induced by vertices $\{v_1, v_6, v_7, v_{16}\}$ has an influence value $\omega(v_7) = 14$ and a minimum degree at 3. It can also be verified that $v_6$ is not a keynode regarding $\gamma = 3$. In the following, for presentation simplicity we simply call a vertex a keynode without referring to the $\gamma$ value which can be inferred from the context.

In order to efficiently count the number of influential $\gamma$-communities in a graph, we prove the following lemmas regarding keynode.

Lemma 3.3: *Given a graph $G$ and a value $\tau$, there is at most one influential $\gamma$-community in $G$ with influence value $\tau$.*

Proof: We prove the lemma by contradiction. Assume there are two influential $\gamma$-communities in $G$ with influence value $\tau$, let them be $g_1$ and $g_2$. Then, there is a unique vertex $u$ in $G$ with $\omega(u) = \tau$; moreover, $u$ is in both $g_1$ and $g_2$. Note that, according to the definition of influential $\gamma$-community, none of $g_1$ and $g_2$ can be a proper subgraph of the other. Let $V_1$ and $V_2$ be the sets of vertices in $g_1$ and in $g_2$, respectively. It is easy to see that the subgraph $G[V_1 \cup V_2]$ also satisfies the connectivity and cohesiveness constraints of Definition 2.2 and moreover, $G[V_1 \cup V_2]$ is a proper supergraph of both $g_1$ and $g_2$. This contradicts that $g_1$ and $g_2$ are influential $\gamma$-communities. Thus, the lemma holds. □

Lemma 3.4: *There is a one-to-one correspondence between influential $\gamma$-communities in a graph $G$ and keynodes in $G$. Thus, the number of keynodes in $G$ equals the number of influential $\gamma$-communities in $G$.*

Proof: ($\Longrightarrow$) Given an influential $\gamma$-community $g$ in $G$, let $u$ be the vertex in $g$ with the minimum weight. It is easy to see that $u$ is a keynode; we say that $u$ is the corresponding keynode of $g$. According to Lemma 3.3, different influential $\gamma$-communities have different influence values, and thus have different corresponding keynodes.

($\Longleftarrow$) Given a keynode $u$, there is a subgraph $g$ of $G$ such that $g$ has an influence value $\omega(u)$ and the minimum vertex degree of $g$ is at least $\gamma$, according to the definition of keynode. Without loss of generality, assume $g$ is connected. Thus, either $g$ is an influential $\gamma$-community or there is a maximal supergraph $g'$ of $g$ that also has an influence value $\omega(u)$ and minimum vertex degree at least $\gamma$; in the latter case, $g'$ is an influential $\gamma$-community. In either case, there is an influential $\gamma$-community with influence value $\omega(u)$. Moreover, according to Lemma 3.3, this influential $\gamma$-community is unique.

Thus, the lemma holds. □

In the following, given an influential $\gamma$-community $g$, we use key($g$) to denote the unique corresponding keynode in $g$, according to Lemma 3.4; that is, the vertex in $g$ with the minimum weight. Note that, an influential $\gamma$-community may contain multiple keynodes, but it is uniquely determined by the keynode with the smallest weight (*i.e.*, key($g$)). For example, $v_{11}$, $v_7$, $v_{13}$ and $v_5$ are keynodes for the graph in Figure 3 with $\gamma = 3$, and they correspond to the four influential $\gamma$-communities shown in Figures 5(a), 5(b), 5(c), and 5(d), respectively.

**The Algorithm CountIC.** Following Lemma 3.4, we count the number of influential $\gamma$-communities in a graph by computing the set of keynodes in the graph. The pseudocode is shown in Algorithm 2. Given a graph $g$, we first reduce $g$ to its $\gamma$-core (Line 1), which is the maximal subgraph with minimum degree at least $\gamma$ [34], and initialize a sequence keys of keynodes and a sequence cvs of vertices to be empty (Lines 2–3). cvs will be used in influential $\gamma$-community enumeration and will be discussed in Section 3.2.2; we ignore cvs for the current being. Then, while the graph $g$ is not empty (Line 4), we get the vertex $u$ with the minimum weight in $g$ (Line 5), which is a keynode (Line 6), and then remove the keynode $u$ from $g$ and reduce the

**Algorithm 2:** CountIC

**Input**: A graph $g$ and an integer $\gamma$

**Output**: The number of influential $\gamma$-communities in $g$

1 $g \leftarrow$ compute the $\gamma$-core of $g$;
2 keys $\leftarrow \emptyset$;
3 cvs $\leftarrow \emptyset$;
4 **while** $g \neq \emptyset$ **do**
5 $\quad u \leftarrow \arg\min_{v \in g} \omega(v)$;
6 $\quad$ Append $u$ to the end of keys;
7 $\quad$ Remove($u, g$, cvs); /* Compute the $\gamma$-core of $g \backslash u$ */;
8 **return** |keys|;

**Procedure** Remove($u, g$, cvs)
9 Initialize a queue $Q$ by $u$;
10 **while** $Q \neq \emptyset$ **do**
11 $\quad$ Pop a vertex $v$ from $Q$;
12 $\quad$ **for each** *neighbor* $v'$ *of* $v$ *in* $g$ **do**
13 $\quad\quad$ **if** *the degree of* $v'$ *in* $g$ *is* $\gamma$ **then**
14 $\quad\quad\quad$ Push $v'$ into $Q$;
15 $\quad$ Remove $v$ from $g$ and append $v$ to the end of cvs;

resulting graph to its $\gamma$-core (Line 7).

The procedure Remove computes the $\gamma$-core of $g \backslash u$ (*i.e.*, the resulting graph by removing $u$ from $g$). Note that, as input to Remove, the graph $g$ itself is a $\gamma$-core, but $g \backslash u$ may not be. Thus, we only need to invoke Remove for $u$, which will then recursively remove all vertices whose degrees become less than $\gamma$ as a result of removing vertices. This is achieved by the queue $Q$ and checking that the degree of vertex $v'$ before removing $v$ is $\gamma$ (Line 13); thus, each vertex is pushed into the queue $Q$ at most once.

In Algorithm 2, we omit the details of computing $\gamma$-core of $g$ at Line 1. This actually can be achieved by invoking the procedure Remove for each vertex in $g$ whose degree is smaller than $\gamma$.

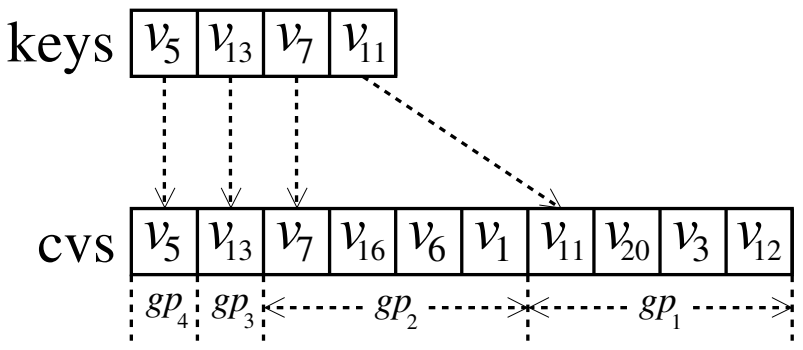

Figure 6: Running example of CountIC

Example 3.2: Consider running CountIC on the subgraph $G_{\geq \tau_2}$ shown in Figure 4(c) for $\gamma = 3$. Initially, we reduce the subgraph to its $\gamma$-core, which removes vertices $\{v_9, v_{17}, v_{18}\}$. Then, we iteratively pick the vertex $u$ with the minimum weight from the remaining graph, add $u$ to keys, and remove $u$ from the graph and also maintain the $\gamma$-core. Firstly, we add $v_5$ to keys, whose removal does not make other vertices' degrees to be smaller than $\gamma$; thus, the procedure Remove merely removes $v_5$ from the graph. Secondly, we similarly add $v_{13}$ to keys and remove it from the graph. Thirdly, we add $v_7$ to keys and remove it from the graph. The removal of $v_7$ makes the degrees of $v_1$, $v_6$, $v_{16}$ become smaller than than $\gamma$; thus, they are all removed from the graph. Similarly, in the fourth step, we add $v_{11}$ to keys and remove all remaining vertices from the graph; the algorithm terminates. The results of keys and cvs are shown in Figure 6. As there are four vertices in keys, we conclude that there are four influential $\gamma$-communities in $G_{\geq \tau_2}$. □

**Time Complexity and Correctness of** CountIC**.** It is easy to see that the time complexity of CountIC (*i.e.*, Algorithm 2) is linear to size of the input graph $g$ (*i.e.*, size($g$)). Note that at Line 13, rather than online counting the degree of a vertex, we maintain the degrees of all vertices at the beginning of Algorithm 2 and also when a vertex is removed from the graph at Line 15. We prove the correctness of Algorithm 2 in the following lemma and theorem.

Lemma 3.5: *After running Algorithm 2,* keys *is the set of keynodes in g.*

Proof: It is easy to verify that every vertex in keys is a keynode. Consider $u$ obtained at Line 5, let $g_u$ be the graph from which $u$ is obtained at Line 5. Then, the connected component of $g_u$ containing $u$ has an influence value $\omega(u)$ and has a minimum degree at least $\gamma$. Thus, $u$ is a keynode.

Now, we prove that every vertex of $g$ that is not in keys is not a keynode. Firstly, it is obvious that the vertices removed at Line 1 when computing the $\gamma$-core of $g$ are not keynodes. Secondly, it is easy to verify by induction that the vertices, other than $u$, removed at Line 15 are also not keynodes. Thus, the lemma holds. □

Theorem 3.2: *Algorithm 2 correctly computes the number of influential $\gamma$-communities in a graph g.*

Proof: This directly follows from Lemmas 3.4 and 3.5. □

### 3.2.2 Influential $\gamma$-community Enumeration

In this subsection, we show that the influential $\gamma$-communities can be obtained from the two arrays, keys and cvs, that are computed by CountIC.

**From** cvs **to Communities.** We call the vertex sequence in cvs as c*ommunity-aware* v*ertex* s*equence*, since the influential $\gamma$-communities can be extracted from it. From Section 3.2.1, we know that each keynode $u$ corresponds to an influential $\gamma$-community with influence value $\omega(u)$, denoted by IC($u$). It is easy to verify that IC($u$) for the $k$ vertices in keys with the largest weights (*i.e.*, the last $k$ vertices) are the top-$k$ influential $\gamma$-communities; note that, vertices in keys are in increasing weight order. In the following, we show how to construct IC($u$) efficiently from keys and cvs.

Firstly, given keys and cvs, we construct one group for each keynode in keys. Denote the group of keynode $u$ by gp($u$), which consists of $u$ and all vertices after $u$ and before the next keynode in cvs; note that all keynodes of keys are in cvs. For example, for the keys and cvs in Figure 6, gp($v_5$) = $\{v_5\}$, gp($v_{13}$) = $\{v_{13}\}$, gp($v_7$) = $\{v_7, v_{16}, v_6, v_1\}$ and gp($v_{11}$) = $\{v_{11}, v_{20}, v_3, v_{12}\}$, where the groups are also shown at the bottom of Figure 6.

Secondly, IC($u$) can be obtained from gp($u$) recursively by the following lemma.

Lemma 3.6: $\mathtt{IC}(u)$ *equals the union of* $\mathtt{gp}(u)$ *and* $\mathtt{IC}(u')$ *for each keynode* $u'$ *(in* $\mathtt{keys}$*) such that* $\omega(u') > \omega(u)$ *and there is an edge between a vertex of* $\mathtt{gp}(u)$ *and a vertex of* $\mathtt{IC}(u')$*; that is,*

$$\mathtt{IC}(u) = \mathtt{gp}(u) \cup \Big( \bigcup_{u' \in \mathtt{keys}, \omega(u') > \omega(u), (\mathtt{gp}(u) \times \mathtt{IC}(u')) \cap E \neq \emptyset} \mathtt{IC}(u') \Big).$$

Proof: Firstly, it is easy to verify that $\mathtt{gp}(u) \subseteq \mathtt{IC}(u)$ for every keynode in $\mathtt{keys}$. Secondly, for any two keynodes $u$ and $v$, either $\mathtt{IC}(v) \cap \mathtt{IC}(u) = \emptyset$ or $\mathtt{IC}(v) \subset \mathtt{IC}(u)$. Moreover, $\mathtt{IC}(v) \subset \mathtt{IC}(u)$ holds if and only if 1) $\omega(v) > \omega(u)$ and there is an edge between $\mathtt{IC}(v)$ and $\mathtt{gp}(u)$, or 2) there is another keynode $v'$ such that $\omega(v') > \omega(u)$ and there is an edge between $\mathtt{IC}(v')$ and $\mathtt{gp}(u)$, and $\mathtt{IC}(v) \subset \mathtt{IC}(v')$; note that, in the latter case, we also have $\mathtt{IC}(v') \subset \mathtt{IC}(u)$. Thus, the lemma holds. □

**Algorithm 3:** EnumIC

**Input**: A graph $g$, a sequence $\mathtt{keys}$ of keynodes, a sequence $\mathtt{cvs}$ of vertices, and an integer $k$
**Output**: Top-$k$ influential $\gamma$-communities

1 $\mathtt{keys} \leftarrow$ the last $k$ keynodes in $\mathtt{keys}$;
2 Initialize $\mathtt{v2key}(v) \leftarrow \mathtt{null}$ for each vertex $v$ in $g$;
3 **for each** *keynode $u$ in* $\mathtt{keys}$ *in reverse order* **do**
4     Initialize $\mathtt{Ch}(u) \leftarrow \emptyset$ and $\mathtt{gp}(u) \leftarrow \emptyset$;
5     **for each** *vertex $v$ in* $\mathtt{cvs}$ *starting from $u$* **do**
6         **if** *$v$ is a keynode **and** $v \neq u$* **then break;**
7         $\mathtt{gp}(u) \leftarrow \mathtt{gp}(u) \cup \{v\}$;
8         $\mathtt{v2key}(v) \leftarrow u$;
9     **for each** *vertex $v$ in $\mathtt{gp}(u)$* **do**
10         **for each** *neighbor $w$ of $v$ in $g$* **do**
11             **if** $\mathtt{v2key}(w) \neq$ null ***and*** $\mathsf{Find}(w, \mathtt{v2key}(\cdot)) \neq u$ **then**
12                 $\mathtt{Ch}(u) \leftarrow \mathtt{Ch}(u) \cup \{\mathsf{Find}(w, \mathtt{v2key}(\cdot))\}$;
13                 $\mathsf{Union}(w, u)$;
14     $\mathtt{IC}(u) \leftarrow \mathtt{gp}(u) \cup \big( \bigcup_{v \in \mathtt{Ch}(u)} \mathtt{IC}(v) \big)$;

**The Algorithm** EnumIC**.** Based on the above discussions, the pseudocode of influential $\gamma$-community enumeration algorithm is shown in Algorithm 3. Firstly, we reduce $\mathtt{keys}$ to contain only the last $k$ vertices (Line 1), and initialize a *disjoint-set data structure* $\mathtt{v2key}$ (Line 2), which maintains for each vertex $v$ the smallest keynode whose corresponding influential $\gamma$-community contains $v$. Then, we process keynodes in $\mathtt{keys}$ in decreasing weight order (Lines 3–14). For each keynode $u$, we firstly obtain the group $\mathtt{gp}(u)$ (Line 7) and initialize $\mathtt{v2key}(v)$ to be $u$ for each $v \in \mathtt{gp}(u)$ (Line 8), and then process the neighbors of vertices in $\mathtt{gp}(u)$ (Lines 9–13). For each neighbor $w$, we add the current smallest keynode whose corresponding influential $\gamma$-community contains $w$ into $\mathtt{Ch}(u)$ (Line 12), and then set $\mathtt{v2key}(\cdot)$ to be $u$ for all vertices in this influential $\gamma$-community (Line 13). Then, we have $\mathtt{IC}(u) = \mathtt{gp}(u) \cup \big( \bigcup_{v \in \mathtt{Ch}(u)} \mathtt{IC}(v) \big)$.

Example 3.3: Consider the $\mathtt{keys}$ and $\mathtt{cvs}$ shown in Figure 6. The 4 keynodes in increasing weight order are $v_5, v_{13}, v_7, v_{11}$. Firstly, we have $\mathtt{gp}(v_{11}) = \{v_{11}, v_{20}, v_3, v_{12}\}$ and $\mathtt{Ch}(v_{11}) = \emptyset$, and $\mathtt{gp}(v_7) = \{v_7, v_{16}, v_6, v_1\}$ and $\mathtt{Ch}(v_7) = \emptyset$; thus, $\mathtt{IC}(v_{11}) = \mathtt{gp}(v_{11})$ and $\mathtt{IC}(v_7) = \mathtt{gp}(v_7)$. Secondly, we have $\mathtt{gp}(v_{13}) = \{v_{13}\}$ and $\mathtt{Ch}(v_{13}) = \{v_{11}\}$, since $v_{13}$ is connected to $v_3, v_{12}, v_{20}$ that are contained in $\mathtt{IC}(v_{11})$; that is, $\mathtt{v2key}(v_3) = \mathtt{v2key}(v_{12}) = \mathtt{v2key}(v_{20}) = v_{11}$. Thus, $\mathtt{IC}(v_{13}) = \mathtt{gp}(v_{13}) \cup \mathtt{IC}(v_{11})$. Similarly, we have $\mathtt{IC}(v_5) = \mathtt{gp}(v_5) \cup \mathtt{IC}(v_7)$. □

**Analysis.** The correctness of Algorithm 3 follows from Lemma 3.6. The time complexity of Algorithm 3 is $O(\mathtt{size}(g))$ by using the technique in [5], where $\mathsf{Find}(\cdot,\cdot)$ and $\mathsf{Union}(\cdot,\cdot)$ are the two fundamental operations on disjoint-set data structure and can be implemented to run in constant amortized time [12]. It is worth noting that, at Line 14, we only link $\mathtt{IC}(v)$ to $\mathtt{IC}(u)$ without actually copying the content of $\mathtt{IC}(v)$ to $\mathtt{IC}(u)$; otherwise, the time complexity is also linear to the output size which can be larger than $\mathtt{size}(g)$.

## 3.3 Analysis of LocalSearch

In the following, we analyze the time complexity of our local search algorithm LocalSearch, discuss the setting of an appropriate $\delta$ value, and prove the instance-optimality of LocalSearch.

**Time Complexity.** Let $\tau^*$ be the target value as defined in Theorem 3.1, and $G_{\geq \tau_h}$ be the subgraph that LocalSearch (*i.e.*, Algorithm 1) accesses before terminating. We prove the time complexity of LocalSearch by the following lemmas and theorem. Recall that, $\delta > 1$ is a parameter used at Line 4 of Algorithm 1.

Lemma 3.7: *The time complexity of* LocalSearch *is* $O((1 + \frac{1}{\delta - 1}) \cdot \mathtt{size}(G_{\geq \tau_h}))$.

Proof: In Algorithm 1, a series of subgraphs (*i.e.*, $G_{\geq \tau_1}, \ldots, G_{\geq \tau_h}$) are constructed and used as input to CountIC for counting influential $\gamma$-communities, and the last subgraph $G_{\geq \tau_h}$ is utilized as input to EnumIC for computing the top-$k$ influential $\gamma$-communities. Note that, each subgraph $G_{\geq \tau}$ can be extracted from $G$ in $O(\mathtt{size}(G_{\geq \tau}))$ time. Thus, the time complexity of LocalSearch is $(\sum_{i=1}^{h} T_1(G_{\geq \tau_i})) + T_2(G_{\geq \tau_h})$, where $T_1(g)$ and $T_2(g)$ represent the time complexities of CountIC and EnumIC, respectively. As $T_1(g) = T_2(g) = O(\mathtt{size}(g))$ and $\mathtt{size}(G_{\geq \tau_i}) \leq \frac{1}{\delta} \mathtt{size}(G_{\geq \tau_{i+1}})$ for $i < h$, the time complexity of LocalSearch is $O(\sum_{i=1}^{h} T_1(G_{\geq \tau_i}) + T_2(G_{\geq \tau_h})) = O(\sum_{i=1}^{h} \mathtt{size}(G_{\geq \tau_i}) + \mathtt{size}(G_{\geq \tau_h})) = O(\sum_{i=1}^{h} \frac{1}{\delta^{h-i}} \mathtt{size}(G_{\geq \tau_h})) = O((1 + \frac{1}{\delta - 1}) \cdot \mathtt{size}(G_{\geq \tau_h}))$. □

Lemma 3.8: *We have* $\mathtt{size}(G_{\geq \tau_h}) < 2\delta \cdot \mathtt{size}(G_{\geq \tau^*})$.

Proof: It is easy to see that $\tau_{h-1} > \tau^* \geq \tau_h$ and $\mathtt{size}(G_{\geq \tau_{h-1}}) < \mathtt{size}(G_{\geq \tau^*})$. Let $u$ be the vertex with the smallest weight in $G_{\geq \tau_h}$ and let $G_{\geq \tau_h} \backslash u$ be the resulting graph of removing $u$ and all its adjacent edges from $G_{\geq \tau_h}$. Then, $\mathtt{size}(G_{\geq \tau_h} \backslash u) < \delta \cdot \mathtt{size}(G_{\geq \tau_{h-1}})$. Moreover, we have $\mathtt{size}(G_{\geq \tau_h}) \leq 2 \cdot \mathtt{size}(G_{\geq \tau_h} \backslash u) + 1$. Thus, $\mathtt{size}(G_{\geq \tau_h}) < 2\delta \cdot \mathtt{size}(G_{\geq \tau^*})$, and the lemma holds. □

Theorem 3.3: *The time complexity of* LocalSearch *is* $O(\frac{2\delta^2}{\delta - 1} \cdot \mathtt{size}(G_{\geq \tau^*}))$.

Proof: This directly follows from Lemmas 3.7 and 3.8. □

**Setting $\delta$.** Following Theorem 3.3, the time complexity of LocalSearch is $O(\mathtt{size}(G_{\geq \tau^*}))$ for any given constant $\delta > 1$. However, the constant factor in the time complexity will be different for different values of $\delta$. In this paper, we set $\delta$ as 2, since $\frac{2\delta^2}{\delta-1}$ achieves the smallest value at $\delta = 2$, among all $\delta$ values larger than 1; note that, $\frac{2\delta^2}{\delta-1} = 2(1 + \delta + \frac{1}{\delta-1})$.

**Instance-optimality of LocalSearch.** Let $\mathcal{A}$ be the class of algorithms that correctly compute top-$k$ influential communities *without indexes* and knowing only the vertex weight vector of the graph $G$, while all other information (such as degree/neighbors of a vertex) are obtained through accessing edges of the graph; that is, obtaining the degree of any vertex in $G_{\geq\tau}$ takes linear time to its number of neighbors in $G_{\geq\tau}$. Then, LocalSearch is a member of $\mathcal{A}$. We prove that LocalSearch is instance-optimal [13] within the class $\mathcal{A}$ of algorithms, by the following lemma and theorem.

Lemma 3.9: *Given a graph G, any algorithm in $\mathcal{A}$ needs to access a subgraph of G of size* $\Omega(\texttt{size}(G_{\geq\tau^*}))$.

Proof: Let $n$ be the number of vertices of $G_{\geq\tau^*}$, we prove that any algorithm of $\mathcal{A}$ needs to know the degrees (and thus all neighbors) of at least $n - \gamma$ vertices of $G_{\geq\tau^*}$. Let's consider an arbitrary algorithm $A$ that computes the top-$k$ influential communities by accessing the full lists of neighbors of only $n-\gamma-1$ vertices. Let $S$ be the set of $\gamma + 1$ vertices whose lists of neighbors are not accessed in full, and $\tau_S$ be the minimum vertex weight of $S$. Then, (1) we have $\tau_S > \tau^*$ according to the definition of $\tau^*$ in Theorem 3.1 and the assumption that each vertex has a distinct weight (see Section 2), and (2) the reported top-$k$ influential communities cannot contain any vertex of $S$ since we need to report all the edges of each community. However, $S$ itself may form a clique in $G_{\geq\tau^*}$ such that there is an influential $\gamma$-community containing $S$ with influence value $\tau_S$, which is larger than the influence value $\tau^*$ of one of the $k$ reported influential $\gamma$-communities; we cannot exclude this possibility without accessing $S$ and without indexes. As a result, algorithm $A$ is incorrect and does not belong to $\mathcal{A}$.

Consequently, for any algorithm $B$ in $\mathcal{A}$, the number of edges of $G_{\geq\tau^*}$ that are not accessed by $B$ is at most $k^2$, which is smaller than $\frac{1}{2}\texttt{size}(G_{\geq\tau^*})$ since the number of edges in an influential $\gamma$-community is at least $k \cdot (k + 1)$. Thus, $B$ needs to access a subgraph of $G$ of size $\Omega(\texttt{size}(G_{\geq\tau^*}))$. □

Note that, Lemma 3.9 is for the case that each vertex has a distinct weight. This lemma also holds if the number of same-weight vertices is bounded by a constant. This is because the proof of Lemma 3.9 essentially implies the number of unvisited vertices in $G_{\geq\tau^*}$ with weight larger than $\tau^*$ is bounded by $\gamma + 1$.

Theorem 3.4: LocalSearch *is instance-optimal within the class $\mathcal{A}$ of algorithms.*

Proof: This follows from Theorem 3.3 and Lemma 3.9. □

**Remarks.** Note that, the time complexity and the instance-optimality of LocalSearch in above are analyzed based on the assumption that the set $N_G(u)$ of neighbors of each vertex is pre-partitioned into two disjoint sets, $N_G^{\geq}(u)$ and $N_G^{<}(u)$ (see Section 3.1), such that any subgraph $G_{\geq\tau}$ can be extracted in $O(\texttt{size}(G_{\geq\tau}))$ time. If this assumption does not hold, then we need to revise the definition of $G_{\geq\tau}$ to be consisting of all the adjacent edges in $G$ for every vertex of $V_{\geq\tau}$. Nevertheless, the time complexity and instance-optimality of LocalSearch still hold based on the revised definition of $G_{\geq\tau}$, by using the same arguments as above.

In Algorithm 1, we choose to grow the subgraph $G_{\geq\tau_i}$ exponentially, based on which we prove the instance-optimality of LocalSearch in above. Another natural choice of growing $G_{\geq\tau_i}$ is that $\texttt{size}(G_{\geq\tau_i}) = i \cdot m$ for a constant $m$; that is, add an additional total $m$ vertices and edges to the subgraph each time. However, then the time complexity would be $(\sum_{i=1}^{h} T_1(G_{\geq\tau_i})) + T_2(G_{\geq\tau_h}) = h^2 \cdot m$ which is super-linear (or even quadratic when $h \gg m$) to the size of the subgraph $G_{\geq\tau_h}$ accessed by the algorithm, as $\texttt{size}(G_{\geq\tau_h}) = h \cdot m$. This validates our choice of exponentially growing $G_{\geq\tau_i}$.

# 4 A Progressive Approach

In Algorithm 1, as well as in existing global search algorithms in [8, 26, 27], the influential $\gamma$-communities are only constructed and reported at the end of an algorithm; that is, the results are only available to the user when the algorithm terminates. Thus, there is a long latency delay between issuing a query and seeing any result. In this section, we propose techniques to compute and report the influential $\gamma$-communities progressively in decreasing influence value order. As a by-product of our progressive approach, the user no longer needs to specify $k$ in the query, and can terminate the algorithm once having seen enough results.

**A Progressive Framework.** Recall that, Algorithm 1 firstly invokes CountIC on a series of subgraphs (*i.e.*, $G_{\geq\tau_1}, \ldots, G_{\geq\tau_h}$ with $\tau_1 > \cdots > \tau_h$) to determine the proper subgraph for processing, and then invokes EnumIC on the last subgraph $G_{\geq\tau_h}$ to compute and report the top-$k$ influential $\gamma$-communities. From Lemma 3.1 we know that, for any two values $\tau \leq \tau'$, every influential $\gamma$-community in $G_{\geq\tau'}$ is also an influential $\gamma$-community in $G_{\geq\tau}$. Thus, the influential $\gamma$-communities in $G_{\geq\tau_h}$ can actually be partitioned into influential $\gamma$-communities in $G_{\geq\tau_1}$, and influential $\gamma$-communities in $G_{\geq\tau_i}$ but not in $G_{\geq\tau_{i-1}}$ for every $1 < i \leq h$. As a result, for each $G_{\geq\tau_i}$ with $1 \leq i \leq h$, we can compute and report a set of influential $\gamma$-communities.

**Algorithm 4:** LocalSearch-P

**Input**: A graph $G = (V, E, \omega)$, and an integer $\gamma$
**Output**: Influential $\gamma$-communities in $G$ in decreasing influence value order

$\tau_1 \leftarrow$ the largest $\tau$ value such that $G_{\geq\tau}$ would contain an influential $\gamma$-community;
$\tau_0 = \tau_{max}$; /* $\tau_{max}$ is the largest vertex weight in $G$ */;
$i \leftarrow 1$;
**while** true **do**
  ConstructCVS($G_{\geq\tau_i}, \gamma, \tau_{i-1}$);
  Output influential $\gamma$-communities in EnumIC-P($G_{\geq\tau_i}$, keys, cvs);
  **if** $G_{\geq\tau_i} = G$ **then break**;
  $\tau_{i+1} \leftarrow \max\{\{\tau \mid \texttt{size}(G_{\geq\tau}) \geq 2 \cdot \texttt{size}(G_{\geq\tau_i})\} \cup \{\tau_{min}\}\}$; /* $\tau_{min}$ is the smallest vertex weight in $G$ */;
  $i \leftarrow i + 1$;

Based on the above ideas, our progressive framework is shown in Algorithm 4. We initialize $\tau_1$ be the largest $\tau$ value such that $G_{\geq\tau}$ would contain an influential $\gamma$-community (Line 1), and $\tau_0$ be the largest vertex weight in $G$ (Line 2). Then, we iteratively construct the keys and cvs for $G_{\geq\tau_i}$ (Line 5), compute and report the influential $\gamma$-communities in $G_{\geq\tau_i}$ that are not contained in $G_{\geq\tau_{i-1}}$ (Line 6), find the next largest $\tau_{i+1}$ such that the size of $G_{\geq\tau_{i+1}}$ is at least twice the size of $G_{\geq\tau_i}$ (Line 8), and increment $i$ by 1 (Line 9). Note that, Algorithm 4 is terminated either when all influential $\gamma$-communities in $G$ have been computed (Line 7) or when a user manually terminates it.

**Algorithm 5:** ConstructCVS

**Input**: A graph $g$, an integer $\gamma$, and a threshold $\tau$
**Output**: keys and cvs

```
g ← compute the γ-core of g;
keys ← ∅;
cvs ← ∅;
while g ≠ ∅ do
    u ← arg min_{v∈g} ω(v);
    if ω(u) ≥ τ then break;
    Append u to the end of keys;
    Remove(u, g, cvs);    /* Compute the γ-core of g\u */;
```

**Incrementally Construct cvs.** From Algorithm 2, it can be verified that the keys and cvs constructed for $G_{\geq\tau_i}$ is a suffix of that constructed for $G_{\geq\tau_{i+1}}$. Moreover, given the influential $\gamma$-communities in $G_{\geq\tau_i}$ and to compute the influential $\gamma$-communities that are in $G_{\geq\tau_{i+1}}$ but not in $G_{\geq\tau_i}$, we only need the prefixes of keys and cvs that does not contain any keynodes in $G_{\geq\tau_i}$. Thus, we can incrementally construct keys and cvs for $G_{\geq\tau_{i+1}}$ by terminating the construction once the next keynode belongs to $G_{\geq\tau_i}$. The pseudocode of incrementally constructing cvs is shown in Algorithm 5, which is similar to Algorithm 2. But in Algorithm 5, rather than counting the number of influential $\gamma$-communities in $g$, we construct the parts of keys and cvs that correspond to keynodes with weights smaller than a given threshold $\tau$.

**Incrementally Enumerate Influential $\gamma$-communities.** The pseudocode of incrementally enumerating influential $\gamma$-communities, denoted by EnumIC-P, is similar to Algorithm 3 with the following differences. Firstly, we retain all keynodes in keys; that is, Line 1 of Algorithm 3 is removed. Secondly, the disjoint-set data structure v2key is a global structure that is shared among different runs of EnumIC-P; moreover, the v2key($v$) of $v$ is only lazily initialized for vertices in cvs.

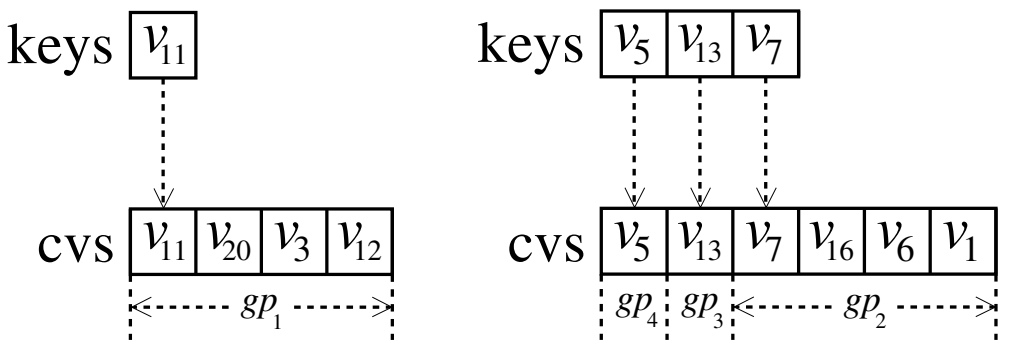

(a) cvs for $G_{\geq\tau_1}$ (b) cvs for $G_{\geq\tau_2}$

Figure 7: Running example of LocalSearch-P

**Running Example of LocalSearch-P.** Consider the graph $G$ in Figure 3 with $\gamma = 3$, and assume the first graph $G_{\geq\tau_1}$ obtained by LocalSearch-P is as shown in Figure 4(b). Firstly, Figure 7(a) shows the keys and cvs computed by ConstructCVS for $G_{\geq\tau_1}$, from which we can obtain the top-1 influential $\gamma$-community as IC($v_{11}$) $= \{v_{11}, v_{20}, v_3, v_{12}\}$. Secondly, Figure 7(b) shows the keys and cvs computed by ConstructCVS for $G_{\geq\tau_2}$, where $v_{11}$ is not included in keys. From the newly constructed keys and cvs, we can obtain the top-2, top-3, and top-4 influential $\gamma$-communities as IC($v_7$), IC($v_{13}$) and IC($v_5$). Moreover, we can see that the concatenation of the two keys in Figure 7 is the same as the keys in Figure 6; this also holds for cvs.

**Time Complexity of LocalSearch-P.** For an arbitrary $k$, let $\tau_k^*$ be the largest value such that $G_{\geq\tau_k^*}$ contains $k$ influential $\gamma$-communities. Then, the time complexity of LocalSearch-P is $O(\texttt{size}(G_{\geq\tau_k^*}))$ whenever a user terminates the algorithm immediately after reporting $k$ influential $\gamma$-communities for an arbitrary $k$. The reasons are the same as in Section 3.3. Thus, the instance-optimality of LocalSearch carries over to LocalSearch-P.

# 5 Extensions

In this section, we extend our framework and techniques to non-containment community search and to other cohesiveness measures.

## 5.1 Non-containment Community Search

According to the definition of influential $\gamma$-community, it is possible that one influential $\gamma$-community is a subgraph of another influential $\gamma$-community. The problem of computing top-$k$ non-containment influential communities is also studied in the literature [8, 26], based on the definition below.

Definition 5.1: [26] Given a graph $G$ and an integer $\gamma$, an influential $\gamma$-community $g$ is a ***non-containment influential $\gamma$-community*** if it satisfies the ***non-containment*** constraint that none of its subgraph is an influential $\gamma$-community.

It is easy to verify that the set of all non-containment influential $\gamma$-communities is disjoint.

**Computing Top-$k$ Non-containment Influential $\gamma$-communities.** Our local search framework in Algorithm 1 can be used to compute the $k$ non-containment influential $\gamma$-communities with the highest influence values, by slightly modifying CountIC (*i.e.*, Algorithm 2) and EnumIC (*i.e.*, Algorithm 3) as follows. Besides keynode, we also define *non-containment keynode* such that there is a one-to-one correspondence between non-containment keynodes and non-containment influential $\gamma$-communities. A keynode $u$ is a non-containment keynode if every vertex that is removed during running the procedure Remove in Algorithm 2 by giving $u$ as input is not connected to any remaining vertex of $g$ obtained after finishing the procedure. Thus, we mark $u$ as a non-containment keynode after Line 7 of Algorithm 2 if this condition holds. Then, the non-containment influential $\gamma$-community corresponding to a non-containment keynode $u$ is exactly $\texttt{gp}(u)$ (see Section 3.2.2 for the definition of $\texttt{gp}(\cdot)$).

Let $\tau^*$ be the largest value such that $G_{\geq\tau^*}$ contains at least $k$ non-containment influential $\gamma$-communities. It can be verified that the time complexity of computing top-$k$ non-containment influential $\gamma$-communities is also $O(\texttt{size}(G_{\geq\tau^*}))$. Nevertheless, this subgraph $G_{\geq\tau^*}$ is no smaller than that for computing top-$k$ influential $\gamma$-communities, due to the fact that the set of all non-containment influential $\gamma$-communities is a subset of all influential $\gamma$-communities. Thus, it is expected that computing top-$k$ non-containment influential $\gamma$-communities takes longer time than computing top-$k$ influential $\gamma$-communities.

## 5.2 Other Cohesiveness Measures

Our framework in Section 3.1 can also be extended to the general case of top-$k$ influential community search regarding other cohesiveness measures. We start with a general definition of influential $\gamma$-cohesive community.

Definition 5.2: Given a vertex-weighted graph $G = (V, E, \omega)$ and a parameter $\gamma$, an ***influential $\gamma$-cohesive community*** is a subgraph $g$ of $G$ such that the following constraints are satisfied.

- ***Connected:*** $g$ is a connected subgraph;
- ***Cohesive:*** the cohesiveness value of $g$ is at least $\gamma$;
- ***Maximal:*** there exists no other subgraph $g'$ of $G$ such that (1) $g'$ is a supergraph of $g$ with $f(g') = f(g)$, and (2) $g'$ is also connected and cohesive.

Note that in the above definition, we do not specify the exact measure of cohesiveness, and it can be any of minimum degree (aka, $k$-core) [32, 34], average degree (aka, edge density) [7, 17], minimum number of triangles each edge participates in (aka, $k$-truss) [11, 31], edge connectivity (aka, $k$-edge connected components) [6, 40], and etc. The influential $\gamma$-community defined in Section 2 is influential $\gamma$-cohesive community where the cohesiveness of a graph is measured by the minimum degree.

**A General Framework for Top-$k$ Influential Community Search.** In order for our framework in Algorithm 1 to be applicable to general top-$k$ influential community search regarding other cohesiveness measures, the influential $\gamma$-cohesive community should satisfy the following two properties.

***Property-I:*** For any two values $\tau_1 \leq \tau_2$, every influential $\gamma$-cohesive community in $G_{\geq\tau_2}$ is also an influential $\gamma$-cohesive community in $G_{\geq\tau_1}$ (similar to Lemma 3.1).

***Property-II:*** For any two values $\tau_1 \leq \tau_2$ and an influential $\gamma$-cohesive community $g$ in $G_{\geq\tau_1}$, if the influence value of $g$ is no smaller than $\tau_2$, then $g$ is also an influential $\gamma$-cohesive community in $G_{\geq\tau_2}$ (similar to Lemma 3.2).

It can be verified that our definition of influential $\gamma$-cohesive community with any of minimum degree, average degree, minimum number of triangles each edge participates in, and edge connectivity, satisfies the above two properties. Thus, we can prove a similar theorem to Theorem 3.1, as follows.

Theorem 5.1: *Let $\tau^*$ be the largest value such that $G_{\geq\tau^*}$ contains at least k influential $\gamma$-cohesive communities. Then, the set of top-k influential $\gamma$-cohesive communities in $G_{\geq\tau^*}$ is the set of top-k influential $\gamma$-cohesive communities in G.*

Proof: This can be proved in a similar way to Theorem 3.1. □

**Algorithm 6:** LocalSearch-General

**Input**: A graph $G = (V, E, \omega)$, and two integers $k$ and $\gamma$
**Output**: Top-$k$ influential $\gamma$-cohesive communities in $G$

```
τ_1 ← the largest τ value such that G_{≥τ} would contain at least k
    influential γ-cohesive communities;
i ← 1;
while CountICC(G_{≥τ_i}, γ) < k and G_{≥τ_i} ≠ G do
    τ_{i+1} ← max {{τ | size(G_{≥τ}) ≥ 2 · size(G_{≥τ_i})} ∪ {τ_min}};
    /* τ_min is the smallest vertex weight in G */;
    i ← i + 1;
return top-k influential γ-cohesive communities in
    EnumICC(G_{≥τ_i});
```

Based on Theorem 5.1, we can easily generalize our local search framework in Algorithm 1 to general top-$k$ influential community search regarding other cohesiveness measures as mentioned above. The pseudocode of our general local search framework is shown in Algorithm 6. CountICC and EnumICC are procedures for counting and enumerating the influential $\gamma$-cohesive communities in a graph, respectively, and only these two procedures need to be specifically designed for different cohesiveness measures.

*Time Complexity.* Let $T_{\mathsf{Count}}(g)$ and $T_{\mathsf{Enum}}(g)$ be the time complexities of CountICC and EnumICC, respectively, for an input graph $g$. The time complexity of Algorithm 6 is as follows.

Theorem 5.2: *If $T_{\mathsf{Count}}$ is linear or super-linear, then the time complexity of Algorithm 6 is $O(T_{\mathsf{Count}}(G_{\geq\tau^*}) + T_{\mathsf{Enum}}(G_{\geq\tau^*}))$, where $\tau^*$ is as defined in Theorem 5.1.*

Proof: This can be proved in a similar way to the proofs of Lemmas 3.7 and 3.8. □

Given a graph $g$, a naive approach to CountICC($g$) for all these cohesiveness measures is iteratively (1) computing the maximal $\gamma$-cohesive subgraph of $g$ and reassigning it as $g$, and (2) removing the minimum-weight vertex from $g$ and marking it as a keynode. This can be optimized by sharing the computation among different iterations (*e.g.*, Algorithm 2); we illustrate the optimized version of CountICC for influential $\gamma$-truss community in below.

**Case Study of Influential $\gamma$-truss Community Search.** Now, we illustrate the procedures CountICC and EnumICC for influential $\gamma$-cohesive community search, where the cohesiveness of a graph is measured by the minimum number of triangles each edge participates in; that is, the cohesiveness of a graph is $\gamma$ if every edge of the graph participates in at least $\gamma - 2$ triangles [21]. We call this definition as influential $\gamma$-truss community.

*The Procedure* CountICC. CountICC counts the number of in-

**Algorithm 7:** CountICC

**Input**: A graph $g$ and an integer $\gamma$
**Output**: The number of influential $\gamma$-truss communities in $g$

```
1 g ← compute the γ-truss of g;
2 keys ← ∅;
3 cvs ← ∅;
4 while g ≠ ∅ do
5     u ← arg min_{v∈g} ω(v);
6     Append u to the end of keys;
      /* Compute the γ-truss of g\u */
7     for each adjacent edge (u, u') of u in g do
8         RemoveEdge((u, u'), g, cvs);
9 return |keys|;
```

fluential $\gamma$-truss community in a graph in a similar way to CountIC in Algorithm 2. That is, it also computes the set of keynodes in the graph, and the number of influential $\gamma$-truss communities in the graph equals the number of keynodes. In addition, it also computes a sequence `cvs` of edges which will be used for enumerating the influential $\gamma$-truss communities. The pseudocode of CountICC is shown in Algorithm 7. It first reduces the graph $g$ to be its $\gamma$-truss (Line 1), where isolated vertices are removed. Then, it iteratively removes the minimum-weight vertex from the graph (Line 5) and reduces the resulting graph to be its $\gamma$-truss (Lines 7–8). Note that, here the truss computation algorithm iteratively removes edges that participate in less than $(\gamma - 2)$-triangles. Thus, `cvs` consists of a sequence of edges.

It is easy to see that the time complexity of CountICC is the same as that of computing $\gamma$-truss of a graph $g$; that is, $O(|E(g)| \cdot \alpha(g))$ [38]. Here, $\alpha(g)$ denotes the arboricity of a graph $g$ which equals the minimum number of forests needed to cover all edges of $g$ [9], and it holds that $\alpha(g) \leq \sqrt{|E(g)|}$.

*The Procedure* EnumICC. Based on the `keys` and `cvs` computed by CountICC, the influential $\gamma$-truss communities in $g$ can also be enumerated in $O(\mathtt{size}(g))$ time, in a similar way to EnumIC in Algorithm 3. Note that, here the time complexity of EnumICC is smaller than that of CountICC. Nevertheless, this does not contradict our earlier claim that counting influential $\gamma$-communities is easier than enumerating. This is because, even if we only want to enumerate the communities without counting them, we still need to first invoke CountICC and then invoke EnumICC; that is, some of the computations of EnumICC are actually done by CountICC.

# 6 Experiments

We conduct extensive performance studies to evaluate the efficiency of our local search framework and algorithms. Firstly, regarding main memory algorithms for influential $\gamma$-community search, we evaluate the following algorithms.

- OnlineAll: the existing global search algorithm in [26].
- Forward: the state-of-the-art global search algorithm in [8].
- Backward: the existing local search algorithm in [8].
- LocalSearch: our *optimal* local search algorithm (Algorithm 1).
- LocalSearch-OA: our local search algorithm by replacing CountIC with OnlineAll.
- LocalSearch-P: our *optimal* and *progressive* local search algorithm (Algorithm 4).

Secondly, we evaluate the following I/O-efficient algorithms.

- OnlineAll-SE: the semi-external version of OnlineAll [27].
- LocalSearch-SE: our semi-external version of LocalSearch-P, where edges are stored on disk (see Remark in Section 3.1).

Thirdly, regarding the extension of our framework to influential $\gamma$-truss community search, we evaluate the following two algorithms.

- LocalSearch-Truss: our local search algorithm for computing top-$k$ influential $\gamma$-truss communities (Algorithm 6).
- GlobalSearch-Truss: a global search algorithm which first invokes CountICC on the entire graph, and then runs EnumICC for enumerating the top-$k$ influential $\gamma$-truss communities.

All algorithms are implemented in C++ and compiled by GNU GCC 4.8.2 with the -O3 flag; the source code of OnlineAll is obtained from the authors of [26] while other algorithms are implemented by us. All experiments are conducted on a machine with an Intel i5 3.20GHz CPU and 16GB main memory.

| Graphs | #vertices | #edges | $d_{max}$ | $d_{avg}$ | $\gamma_{max}$ |
|---|---|---|---|---|---|
| Email | 36,692 | 183,831 | 1,383 | 10.02 | 43 |
| Youtube | 1,134,890 | 2,987,624 | 28,754 | 5.27 | 51 |
| Wiki | 1,791,489 | 25,446,040 | 238,342 | 28.41 | 99 |
| Livejournal | 3,997,962 | 34,681,189 | 14,815 | 17.35 | 360 |
| Orkut | 3,072,627 | 117,185,083 | 33,313 | 76.28 | 253 |
| Arabic | 22,744,080 | 553,903,073 | 575,628 | 48.71 | 3,247 |
| UK | 39,459,925 | 783,027,125 | 1,776,858 | 39.69 | 588 |
| Twitter | 41,652,230 | 1,468,365,182 | 2,997,487 | 70.51 | 2,488 |

Table 1: Statistics of real graphs

**Real Graphs.** We evaluate the algorithms on eight real graphs: Email, Youtube, Wiki, Livejournal, Orkut, Arabic, UK, and Twitter. The first five graphs are downloaded from the Stanford Network Analysis Platform[2], while the last three are downloaded from the Laboratory of Web Algorithmics[3]. Statistics of the graphs are given in Table 1, where $\gamma_{max}$ denotes the maximum value such that the graph contains a non-empty $\gamma_{max}$-core. The weights of vertices are assigned as their PageRank values with the damping factor being set as 0.85.[4]

**Query Parameters.** There are two query parameters, $k$ and $\gamma$. We choose $k$ from $\{5, 10, 20, 50, 100\}$ and $\gamma$ from $\{5, 10, 20, 50\}$; $k = 10$ and $\gamma = 10$ by default. Note that, as $\gamma_{max}$ for Email is 43 as shown in Table 1, the largest $\gamma$ we tested for Email is 40.

[2] `http://snap.stanford.edu/`
[3] `http://law.di.unimi.it/datasets.php`
[4] `https://en.wikipedia.org/wiki/PageRank`

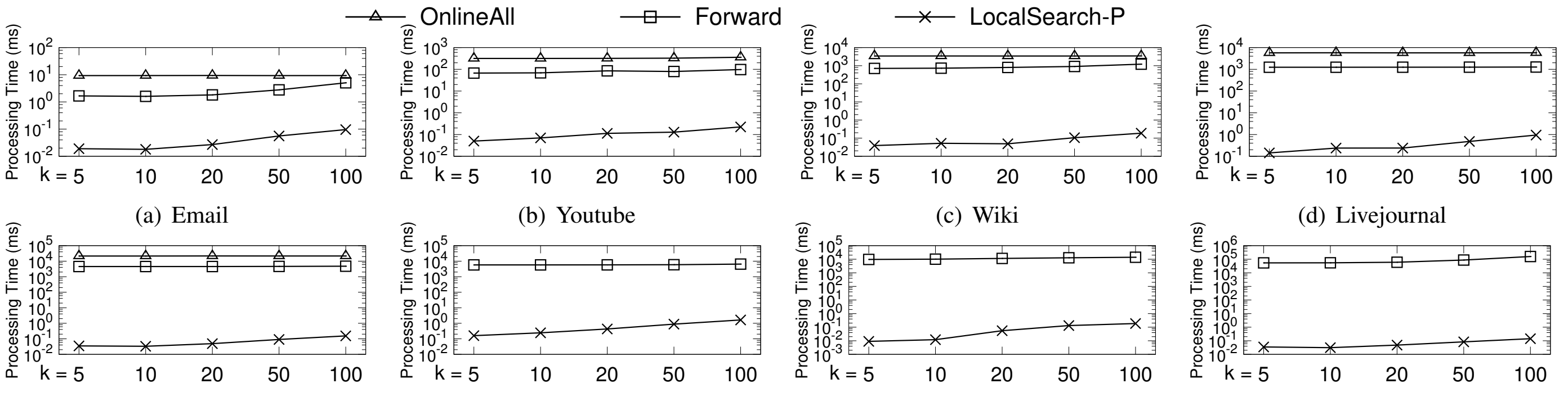

(a) Email (b) Youtube (c) Wiki (d) Livejournal

(e) Orkut (f) Arabic (g) UK (h) Twitter

Figure 8: Against existing global search algorithms ($\gamma$ = 10, vary $k$)

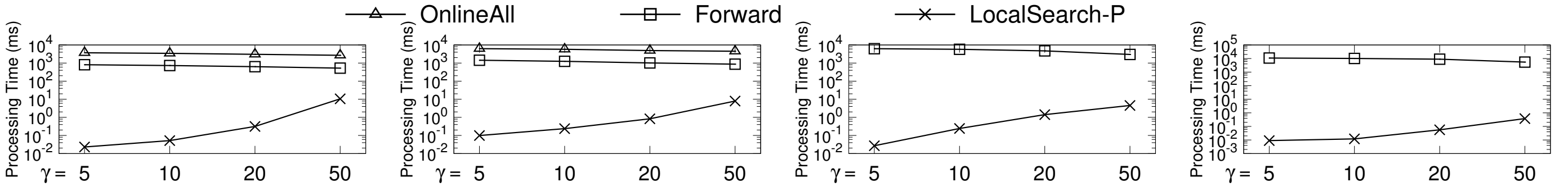

(a) Wiki (b) Livejournal (c) Arabic (d) UK

Figure 9: Against existing global search algorithms ($k$ = 10, vary $\gamma$)

In each testing, for a query with given $k$ and $\gamma$, we run an algorithm on a graph three times and report the average CPU time in milliseconds. For main memory algorithms, the graph is assumed to be stored in main memory, while for I/O-efficient algorithms, the reported time also includes the I/O time.

## 6.1 Experimental Results

**Eval-I: Against Global Search Algorithms by Varying $k$ and $\gamma$.** In this testing, we evaluate LocalSearch-P against the existing global search algorithms OnlineAll and Forward by varying $k$ and $\gamma$. The processing time of the algorithms by varying $k$ is shown in Figure 8, where $\gamma$ = 10. We can see that the processing time of OnlineAll and Forward remains almost the same for different $k$ values. This is because, these two algorithms need to process the entire input graph regardless of the value of $k$. On the other hand, LocalSearch-P runs slower for larger $k$, due to our local search framework that needs to access a larger subgraph for computing more influential $\gamma$-communities. Nevertheless, LocalSearch-P significantly outperforms OnlineAll and Forward across all different $k$ values, and the improvement can be up-to 5 orders of magnitude (*e.g.*, on Orkut). Note that, we omit OnlineAll for Arabic, UK, and Twitter, since it runs out-of-memory for processing these graphs.

The results by varying $\gamma$ are shown in Figure 9, where $k$ = 10. Similar to the results in Figure 8, the processing time of OnlineAll and Forward remains almost the same for different $\gamma$ values. The processing time of LocalSearch-P increases for larger $\gamma$ value. This is because, the larger the value of $\gamma$, the smaller the influence values of the top-$k$ influential $\gamma$-communities. Thus, LocalSearch-P needs to access a larger subgraph for computing top-$k$ influential $\gamma$-communities of larger $\gamma$. Nevertheless, LocalSearch-P outperforms OnlineAll and Forward regarding all different values of $\gamma$.

We also evaluate the algorithms for large values of $k$ and $\gamma$ on the two graphs Arabic and Twitter that have the largest $\gamma_{max}$ values (see Table 1). The results are shown in Figure 10, and the trend is similar to that of Figures 8 and 9. Although LocalSearch-P takes more time when $k$ or $\gamma$ becomes larger, it still outperforms Forward.

**Eval-II: Against Existing Local Search Algorithm Backward.** In this testing, we evaluate our local search algorithm LocalSearch-P against the existing local search algorithm Backward. The results are shown in Figure 11. The processing time of LocalSearch-P and Backward increases for larger $k$, since both algorithms need to access and process a larger subgraph for computing more influential $\gamma$-communities. Nevertheless, LocalSearch-P consistently outperforms Backward. This is because LocalSearch-P has a linear time complexity regarding the subgraph accessed, while Backward has a quadratic time complexity regarding the subgraph accessed [8]. The improvement of LocalSearch-P over Backward is more evident for larger $\gamma$.

**Eval-III: Evaluate LocalSearch-P against LocalSearch-OA.** In this evaluation, we compare LocalSearch-P against its variant that invokes OnlineAll for counting the number of influential $\gamma$-communities in a given graph, denoted LocalSearch-OA. The results in Figure 12 show that LocalSearch outperforms LocalSearch-OA. Thus, we propose a new algorithm CountIC for counting the number of influential $\gamma$-communities in a graph without enumerating them.

**Eval-IV: Evaluate the exponential growth ratio $\delta$.** In this testing, we evaluate the performance of LocalSearch-P for different values of the growth ratio $\delta$, chosen from $\{1.5, 2, 3, 4, 8, 16, 32, 64, 128\}$. The results are shown in Figure 13. Recall from Section 3.3 that, given any constant $\delta$, our algorithm LocalSearch-P runs in linear time to $\texttt{size}(G_{\geq\tau^*})$, and different values of $\delta$ will result into different constant in the time complexity. As a result, the running time of LocalSearch-P for similar values of $\delta$ are similar. In general,

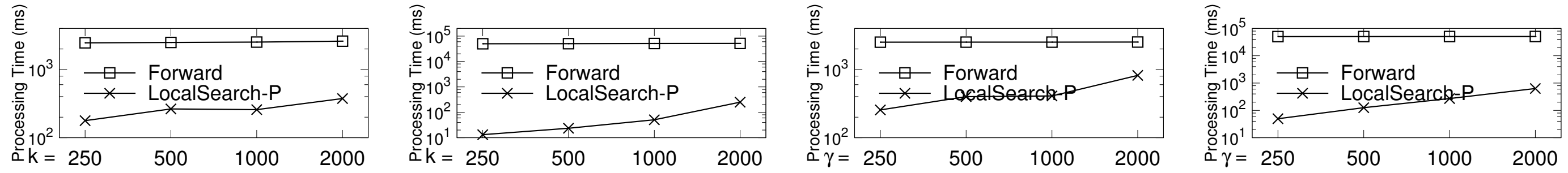

(a) Arabic ($\gamma$ = 1000, vary $k$) (b) Twitter ($\gamma$ = 1000, vary $k$) (c) Arabic ($k$ = 1000, vary $\gamma$) (d) Twitter ($k$ = 1000, vary $\gamma$)

Figure 10: Against Forward for large $k$ and $\gamma$

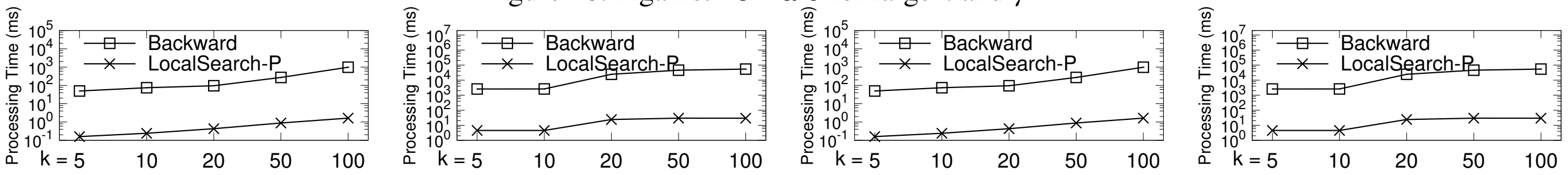

(a) Arabic ($\gamma$ = 10) (b) Arabic ($\gamma$ = 50) (c) UK ($\gamma$ = 10) (d) UK ($\gamma$ = 50)

Figure 11: Against existing local sarch algorithm Backward (vary $k$)

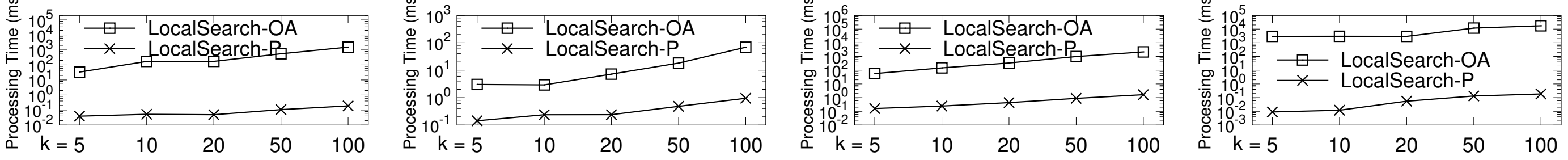

(a) Wiki (b) Livejournal (c) Arabic (d) UK

Figure 12: Evaluate LocalSearch-P against LocalSearch-OA ($\gamma$ = 10, vary $k$)

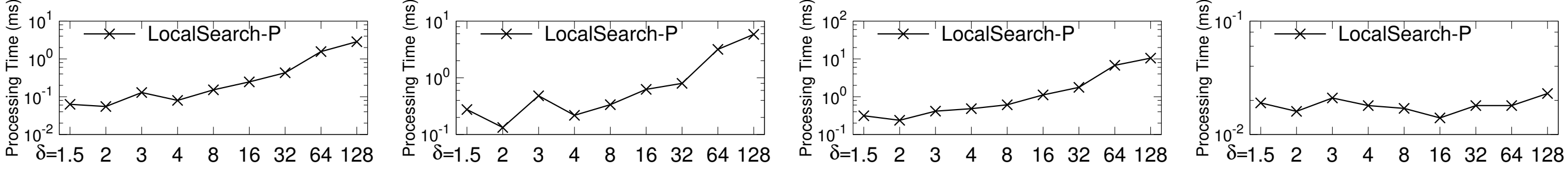

(a) Wiki (b) Livejournal (c) Arabic (d) UK

Figure 13: Evaluate exponential growth ratio $\delta$ ($k$ = 10, $\gamma$ = 10)

the processing time of LocalSearch-P increases for larger $\delta$, and LocalSearch-P performs the best for $\delta$ being around 2.

**Eval-V: Evaluate Our Progressive Approach.** In this testing, we evaluate our progressive approach LocalSearch-P against our non-progressive approach LocalSearch. The experimental results regarding enumeration time are shown in Figure 14. Here $k = 128$, and the numeration time is the elapsed time from the start of the algorithm until the top-$i$ community is reported. As LocalSearch reports the communities one-by-one only at the end of the algorithm, the numeration time for different communities is almost the same. In contrast, LocalSearch progressively computes and reports the communities, and thus the enumeration time increases. As a result, based on our progressive approach LocalSearch-P, the communities are reported to a user progressively as early as possible, and the user can terminate the algorithm once having seen enough communities without the need of specifying $k$ in the query.

The results of evaluating the total processing time of LocalSearch-P and LocalSearch by varying $k$ are shown in Figure 15. We can see that, LocalSearch-P slightly improves upon LocalSearch, despite that LocalSearch-P has the advantage of progressively reporting the communities. This is because LocalSearch-P shares computations among the processing of different subgraphs.

**Eval-VI: Evaluate Our I/O-efficient Algorithm LocalSearch-SE.** In this testing, we evaluate our I/O-efficient algorithm LocalSearch-SE against the exiting one OnlineAll-SE, which is a semi-external version of OnlineAll [27], on two large graphs Arabic and Twitter. OnlineAll-SE iteratively (1) loads as many edges as possible in decreasing weight order from disk to main memory until the memory is full, (2) conducts computation regarding the subgraph in main memory by invoking OnlineAll, and (3) removes from main memory the edges that are already part of communities and thus not needed for the following computations. In this testing, we assume that the main memory can hold $1GB$ of edges in addition to the information regarding vertices. The results of the total processing time are shown in Figure 16. We can clearly see that LocalSearch-SE outperforms OnlineAll-SE, which is a result of our optimal local search framework. Moreover, LocalSearch-SE consumes much smaller main memory compared with OnlineAll-SE, as shown in Figure 17.

**Eval-VII: Evaluate Non-containment Queries.** Here, we evaluate the efficiency of LocalSearch-P for processing non-containment queries, as discussed in Section 5.1; that is, compute the top-$k$ influential $\gamma$-communities such that none of its subgraph is an influential $\gamma$-community [8, 26]. The results of comparing LocalSearch-P with Forward for non-containment queries are shown in Figure 18; note that, here Forward refers to its variant in [8] that computes non-containment communities. We can see that LocalSearch-P clearly outperforms Forward.

**Eval-VIII: Evaluate Influential $\gamma$-truss Community Search Queries.** In this testing, we evaluate the efficiency of our local search approach LocalSearch-Truss for processing influential

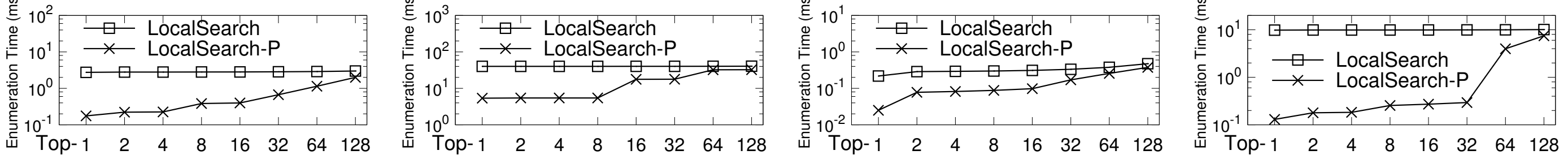

(a) Arabic ($\gamma = 10$) (b) Arabic ($\gamma = 50$) (c) UK ($\gamma = 10$) (d) UK ($\gamma = 50$)

Figure 14: Evaluate our progressive approach regarding enumeration time ($k = 128$)

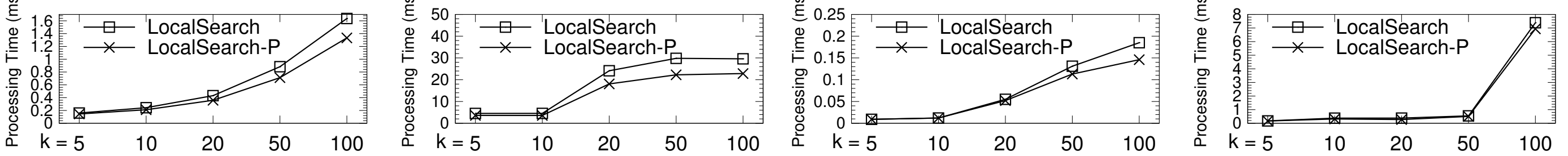

(a) Arabic ($\gamma = 10$) (b) Arabic ($\gamma = 50$) (c) UK ($\gamma = 10$) (d) UK ($\gamma = 50$)

Figure 15: Evaluate our progressive approach regarding total processing time (vary $k$)

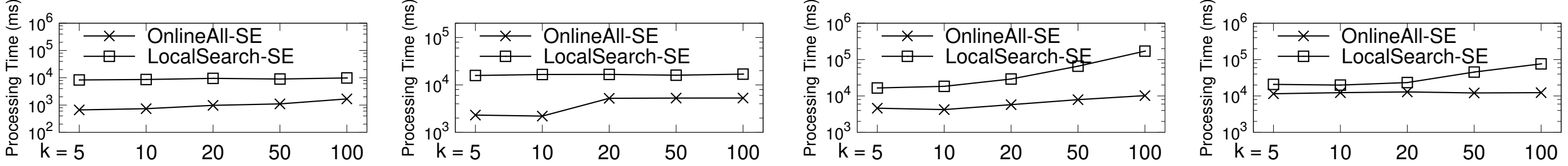

(a) Arabic ($\gamma = 10$) (b) Arabic ($\gamma = 50$) (c) Twitter ($\gamma = 10$) (d) Twitter ($\gamma = 50$)

Figure 16: Evaluate our I/O-efficient algorithm LocalSearch-SE regarding total processing time (vary $k$)

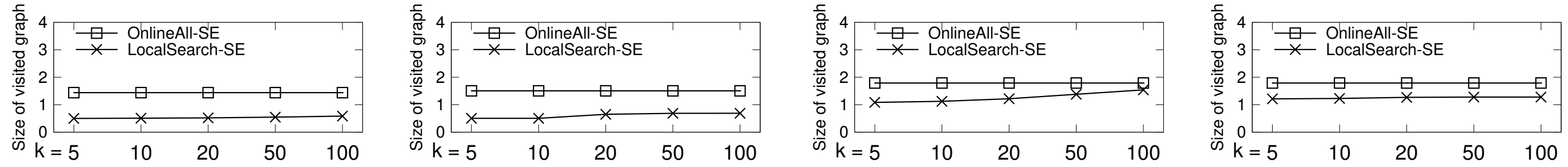

(a) Arabic ($\gamma = 10$) (b) Arabic ($\gamma = 50$) (c) Twitter ($\gamma = 10$) (d) Twitter ($\gamma = 50$)

Figure 17: Evaluate our I/O-efficient algorithm LocalSearch-SE regarding memory usage (vary $k$)

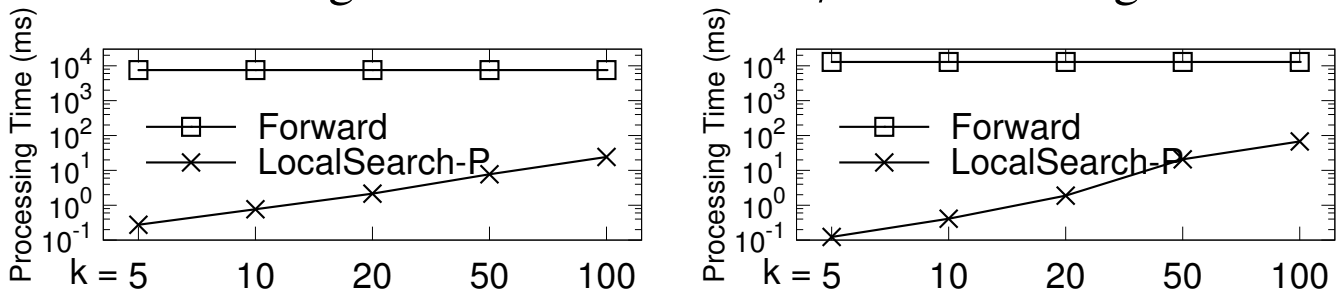

(a) Arabic (b) UK

Figure 18: Evaluate non-containment queries (vary $k$)

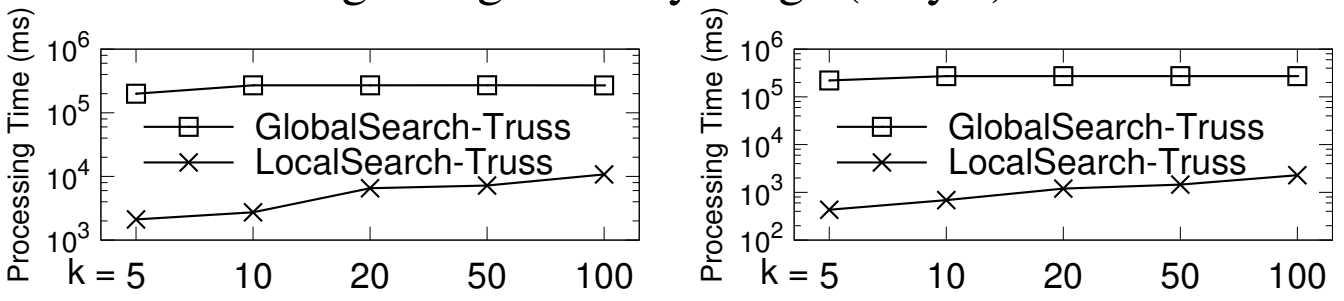

(a) Wiki (b) Livejournal

Figure 19: Evaluate $\gamma$-truss community search queries (vary $k$)

$\gamma$-truss community search queries, by comparing with a global search approach GlobalSearch-Truss that traverses the entire graph. The results are shown in Figure 19, where $\gamma = 10$. We can see that LocalSearch-Truss significantly outperforms GlobalSearch-Truss. This demonstrates the superiority of our local search framework for general top-$k$ influential community search regarding other cohesiveness measures. By comparing Figure 19 with Figure 8, we can see that computing top-$k$ influential $\gamma$-truss communities generally takes more time than computing top-$k$ influential $\gamma$-communities. This is because computing $\gamma$-truss communities has a higher time complexity and also processes a larger subgraph of $G$, than computing $\gamma$-communities.

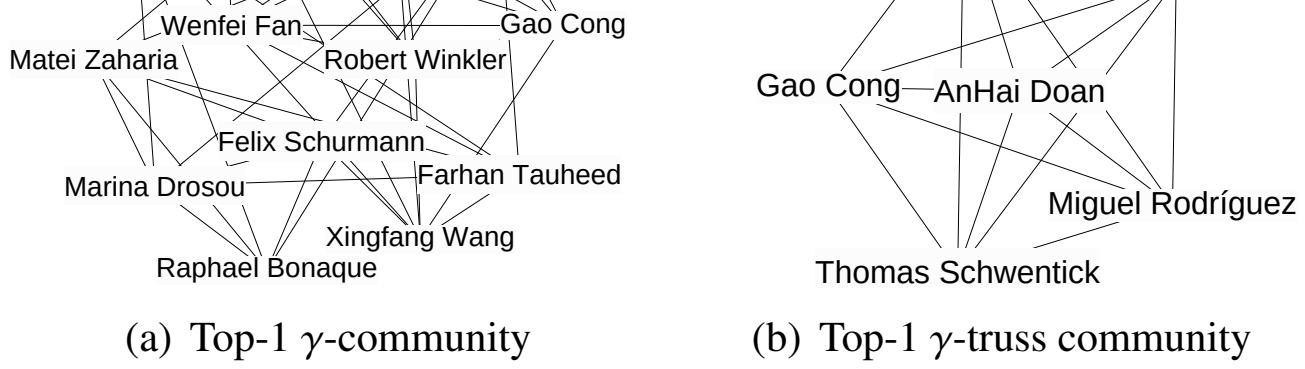

(a) Top-1 $\gamma$-community (b) Top-1 $\gamma$-truss community

Figure 20: Case study on DBLP

**Eval-IX: Case Study on DBLP.** Here, we conduct a case study for the influential $\gamma$-community and $\gamma$-truss community on a co-author network, DBLP. We extract a co-author graph from DBLP (`http://dblp.unitrier.de/xml/`) by focusing on the research areas of *Artificial Intelligence*, *Computer Vision*, *Information Retrieval*, *Data Mining*, *Database*, *Machine Learning* and *Natural Language*. Each vertex corresponds to a researcher that has published at least 10 papers in these research areas, and there is an edge between two researchers if they have co-authored at least 3 papers. Weights of vertices are computed as their PageRank values. The top-1 influential 5-community and 6-truss community are shown in Figures 20(a) and 20(b), respectively; note that, the 5-core community of the vertices in Figure 20(a) consists of $1,148$ vertices, as shown in Figure 21. The minimum weight vertex in Figure 20(a) is "Xingfang Wang" which ranks 215 out of 1743 vertices, and the minimum weight vertex in Figure 20(b) is "AnHai Doan" which ranks 339 out of 1743 vertices; note that, the higher the weight of a vertex the smaller its rank. Thus, although influential $\gamma$-truss community search can find smaller and denser communities, $\gamma$-truss communities usually have smaller influence values than $\gamma$-communities since the $\gamma$-truss constraint is harder to be satisfied than the $\gamma$-core constraint. Note that, for any in-fluential $\gamma$-truss community $g$ with influence value $\tau$, there is a corresponding $(\gamma - 1)$-community with influence value $\tau$ that contains $g$.

# 7 Conclusion

In this paper, we developed a local search framework for the problem of top-$k$ influential community search. We proved that our LocalSearch algorithm for top-$k$ influential $\gamma$-community search is instance-optimal, in the sense that its time complexity is linearly proportional to the size of the smallest subgraph that a correct algorithm needs to access without indexes. We further proposed techniques to make LocalSearch progressively compute and report the influential $\gamma$-communities. We also extended our local search framework to the general case of top-$k$ influential community search regarding other cohesiveness measures. Extensive empirical studies on real graphs demonstrated the superiority of our local search approach over the existing online search algorithms. One direction of future work is to integrate our techniques to the WebGraph framework [3] to process larger graphs in main memory. Another possible direction is extending our techniques to the case that the vertex weight vector is computed online based on the query; for example, the weight of a vertex is computed as the reciprocal of its shortest distance to the query vertices as studied in [23].

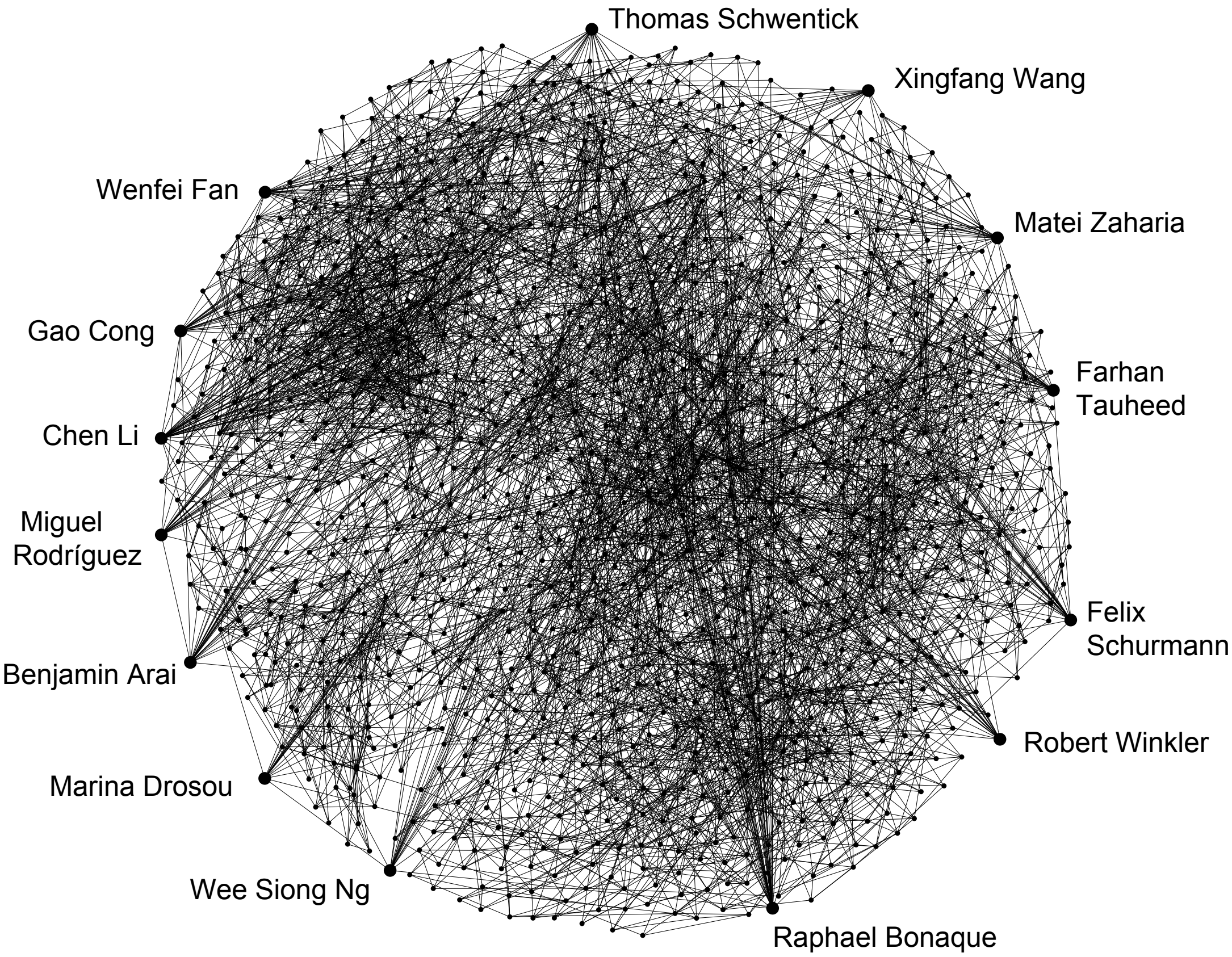

Figure 21: 5-core community of "Xingfang Wang"